\documentclass{LMCS}
\usepackage{enumerate}
\usepackage{hyperref}
\usepackage{Guy,latexsym,amssymb,bbb}
\usepackage{prooftree,neweqnarray}

\newcommand{\pw}{\mathcal{P}\omega}
\newcommand{\graph}[1]{\mathsf{graph}(#1)}
\newcommand{\fun}[1]{\mathsf{fun}(#1)}
\newcommand{\code}{\mathsf{code}}
\newcommand{\retr}{\unlhd}

\newcommand{\sci}{SCI}
\newcommand{\scim}{bSCI}
\newcommand{\scicm}{SCI_{\texttt{mk}, \texttt{ran}}}
\newcommand{\scimk}{SCI_{\texttt{mk}}}

\newcommand{\Comm}{\mathtt{comm}}
\newcommand{\Nat}{\mathtt{nat}}
\newcommand{\Var}{\mathtt{var}}

\newcommand{\while}[2]{\mathtt{while}~#1~\mathtt{do}~#2}
\newcommand{\ifzero}[3]{\mathtt{ifzero}~#1~\mathtt{then}~#2~\mathtt{else}~#3}
\newcommand{\Skip}{\mathtt{skip}}
\newcommand{\assign}{\mathbin{:=}}
\newcommand{\seq}{\mathbin{;}}
\newcommand{\deref}{\mathop{!}}
\newcommand{\new}[2]{\mathtt{new}~#1~\mathtt{in}~#2}

\newcommand{\random}{\mathtt{random}}
\newcommand{\mkvar}[2]{\mathtt{mkvar}~#1~#2}

\newcommand{\override}[3]{(#1 \mid #2 \mapsto #3)}
\newcommand{\cvgs}{\mathord{\Downarrow}}
\newcommand{\eval}{\mathbin{\Downarrow}}
\newcommand{\cequiv}{\cong}
\newcommand{\capprox}{\sqpreord}
\newcommand{\cequivmk}{\cong_{\mathtt{mk}}}
\newcommand{\cequivcm}{\cong_{\mathtt{mk},\mathtt{ran}}}

\newcommand{\naturals}{\bbb{N}}
\newcommand{\readmove}{\mathsf{read}}
\newcommand{\writemove}{\mathsf{write}}
\newcommand{\emptyseq}{\varepsilon}
\newcommand{\strans}[3]{#1 \stackrel{\textstyle{#2}}{\longrightarrow} #3}

\newcommand{\mon}{\mathbf{Mon}}
\newcommand{\monrel}{\mathbf{MonRel}}
\newcommand{\monrelcoh}{\mathbf{MonRelCoh}}
\newcommand{\rel}{\mathbf{Rel}}
\newcommand{\pair}[2]{\langle #1, #2 \rangle}
\newcommand{\powerset}{\mathcal{P}}
\newcommand{\comp}{\mathbin{;}}
\newcommand{\tensor}{\lltensor}
\newcommand{\op}[1]{#1^{\mathrm{op}}}
\newcommand{\ev}{\mathsf{ev}}
\newcommand{\id}{\mathsf{id}}

\newcommand{\produce}[1]{\mathsf{produce}#1}
\newcommand{\test}[1]{\mathsf{test}#1}
\newcommand{\init}[1]{\mathsf{init}(#1)}
\newcommand{\final}[1]{\mathsf{final}{(#1)}}
\newcommand{\rworder}{\preceq}

\newcommand{\coh}{\mathbin{\frown}}

\newcommand{\length}[1]{|#1|}

\def\doi{6 (1:2) 2010}
\lmcsheading%
{\doi}
{1--35}
{}
{}
{May.~14, 2009}
{Jan.~12, 2010}
{}   

\begin{document}

\title{A graph model for imperative computation}
\author{Guy McCusker}
\address{Department of Computer Science\\University of
  Bath\\Bath BA2 7AY\\United Kingdom}
\email{G.A.McCusker@bath.ac.uk}

\begin{abstract}
  Scott's graph model is a
  lambda-algebra based on the observation that continuous
  endofunctions on the lattice of sets of natural numbers can be
  represented via their graphs. A graph is a relation mapping finite
  sets of input values to output values. 
  
  We consider a similar model based on relations whose input values
  are finite sequences rather than sets. This
  alteration means that we are taking into account the order in which
  observations are made. This new notion of graph gives rise to a
  model of affine lambda-calculus that admits an interpretation of
  imperative constructs including variable assignment, dereferencing
  and allocation.
  
  Extending this untyped model, we construct a category that
  provides a model of typed higher-order imperative computation with
  an affine type system. An appropriate language of this kind is
  Reynolds's Syntactic Control of Interference. Our model turns out to be
  fully abstract for this language. At a concrete level, it is
  the same as Reddy's object spaces
  model, which was the first ``state-free''
  model of a higher-order imperative programming language and an
  important precursor of games models. The graph model can therefore
  be seen as a universal domain for Reddy's model. 
\end{abstract}

\maketitle

\section{Introduction}
This paper is an investigation into the semantics of imperative
programs, using a style of model first proposed by
Reddy~\cite{ReddyUS:gloscu}. Reddy's model was a significant
development, because it was the first to model imperative programs
without the use of an explicit semantic entity representing the 
store. Instead, programs are interpreted as ``objects'' (in
Reddy's terminology) which exhibit history-sensitive behaviour. The
store is not modelled explicitly; instead one models the behaviour
that results from the use of the store. 

This new approach turned out to be the key to finding models that are
\emph{fully abstract}: that is, models whose equational theory
coincides with the operationally defined notion of program
equivalence. The first such models for higher-order imperative
programming languages to be discovered were based on game
semantics~\cite{McCuskerGA:linssff,McCuskerGA:fulags}. Although these
models used several ideas from Reddy's work, it was not known whether
Reddy's model was itself fully abstract for the language $\sci$ which
it interprets.

In this paper, some of which is a much extended exposition of work
first presented in~\cite{McCuskerGA:fularm}, we show that Reddy's
model is indeed fully abstract. But more than this, we argue that it
arises from a straightforward modification of Scott's well-known $\pw$
graph-model of the $\lambda$-calculus~\cite{ScottDS:dattl}. Just as in
Scott's work, we develop a model in which every type-object appears as a
retract of a universal object, and it turns out that these retractions are
all definable in a slightly extended \sci\ language. Thus the language
has a \emph{universal type},  which leads to a
very cheap proof of full abstraction. With some additional effort, we
show that the extensions required to establish this universal type are
in fact conservative, that is, they do not alter the notion of program
equivalence. Therefore the original model is itself fully abstract. 

We should remark that the work required to establish conservativity of
one of these extensions amounts to a partial definability result which
would be enough to prove full abstraction of the original model
directly; indeed, that is what was done
in~\cite{McCuskerGA:fularm}. Nevertheless, we believe that the
presentation in terms of conservativity is useful, not least because
of the ease of establishing full abstraction for the extended
language. 

\subsection{Related work}
The utility of a universal type for establishing properties of a model
is well-known, and was explained in detail by
Longley~\cite{LongleyJ:unitwt}. The central idea of this paper, of
modifying Scott's graph model to record slightly different
information, has also been used by Longley in~\cite{LongleyJ:intlce}
to obtain a model of fresh name generation. A similar model
construction has been investigated by Hyland et
al.~\cite{PowerJ:cattfe}. We shall remark further on the connections
between these papers and our present work below, although we leave
closer investigation for future work. 

The denotational semantics of $\sci$ was first treated by
O'Hearn~\cite{OHearnPW:modsci} using functor categories. Reddy's
model~\cite{ReddyUS:gloscu} was the first to avoid the explicit use of
a store-component in the mathematical model, but as mentioned above
this model was not known to be fully abstract until a preliminary
version of the work being reported here
appeared~\cite{McCuskerGA:fularm}. Joint work of the present
author and Wall~\cite{WallM:gamsci} developed a game semantics for
$\sci$ and established a full abstraction
result. Laird~\cite{LairdJ:decscif} analysed the fully abstract relational
model to show that equivalence of programs in a finitary fragment of
$\sci$ is decidable, but observational approximation is not, and went
on to construct a fully abstract games model of a version of $\sci$ with
control operators, establishing decidability of both equivalence and
approximation. The $\sci$ type system itself has been refined and
extended in two ways: first by Reynolds, using intersection
types~\cite{ReynoldsJC:synci2}, and then by O'Hearn et
al.~\cite{OHearnPW:syncir}, using a novel system with two-zone type
judgements. 

\subsection{Acknowledgments}
The author is very grateful to the many researchers with whom he has
discussed this work, including Martin Churchill, Jim Laird, John
Longley, Ana Carolin Martins, Peter O'Hearn, John Power and Uday
Reddy. The comments of anonymous referees were very useful in the
preparation of the final version of the paper. The author also
benefitted from the support of two EPSRC research grants during the
development and preparation of this paper.

\section{Scott's \texorpdfstring{$\pw$}{P-omega}  model}
\label{sec:scott-pw}
We begin with a brief review of Scott's \emph{$\pw$  graph model} of
the $\lambda$-calculus, which appeared in the seminal paper
\emph{Data Types as Lattices}~\cite{ScottDS:dattl}. 

Let $\pw$ denote the lattice of sets of natural numbers, ordered by
inclusion. A continuous function $f: \pw \rightarrow \pw$ is
determined by its action on finite sets. Therefore, such an $f$ is
determined by the set
\[
 \graph{f} = \{ (S, n) \mid S \subseteq_{\mathsf{fin}} \omega, n \in
 \omega, n \in f(S) \}. 
\]
Conversely, let $G$ be a set of pairs $(S, n)$ with $S
\subseteq_\mathsf{fin} \omega$ and $n \in \omega$. We can define 
a continuous function $\fun{G}:\pw \rightarrow \pw$ by 
\[
 \fun{G}(S) = \{ n \mid \exists S' \subseteq S. (S', n) \in G \}
\]
and it is clear that for any continuous $f$, $\fun{\graph{f}} = f$. 

Let $\code(-)$ be any injective encoding 
\[
 \code: \mathcal{P}_\mathsf{fin}\omega \times \omega \rightarrow
 \omega. 
\]
Writing $[\pw \rightarrow \pw]$ for the complete partial order of
continuous functions from $\pw$ to itself, the mapping 
\[
  f \mapsto \{ \code(S,n) \mid (S,n) \in \graph{f}\}
\]
is a continuous function $[\pw \rightarrow \pw] \rightarrow \pw$, and
\[
  S \mapsto \fun{\{(S',n) \mid \code(S',n) \in S \}}
\]
is a continuous function $\pw \rightarrow [\pw \rightarrow
\pw]$. These two mappings therefore form a retraction
\[
[\pw \rightarrow \pw] \retr \pw
\]
in the category of domains and continuous functions, so that $\pw$ is
a reflexive object in this category, and thus a model of untyped
$\lambda$-calculus. For more details on how reflexive objects are
used to model $\lambda$-calculus, see
Barendregt~\cite{BarendregtHP:lamcss}.

Scott in fact worked in the other direction: from the $\pw$ model he
defined a category in which to work, using the \emph{Karoubi envelope}
(see for example~\cite{LambekJ:inthoc})
of the monoid of endomorphisms of $\pw$. One way of presenting this
monoid is as follows. Its elements are graphs of continuous functions 
from $\pw$ to itself; explicitly, an element $a$ is a set of pairs
$(S,n)$, where $S \subseteq_{\mathrm{fin}} \omega$ and $n \in \omega$,
such that \[ (S,n) \in a \land S \subseteq S' \implies (S',n) \in a.\]
(It is easy to verify that these are exactly the image of the $\graph{-}$
function.) 
The monoid operation is the graph representation of function
composition, which can be defined by 
\[
 a \cdot b = \{ (\bigcup_{i=1}^k S_i, n) \mid \exists m_1, \ldots,
 m_k.   (\{m_1, \ldots, m_k\}, n) \in
                 b \land  (S_i, m_i) \in a, i = 1,\ldots, k\}.
\]
The Karoubi envelope of this monoid is the category whose objects are
idempotents, \ie elements $a$ such that $a = a \cdot a$, and maps $f:
a \rightarrow b$ are elements of the monoid such that $f  = a \cdot f
\cdot b$. Scott shows that this is a cartesian closed category and
notes that it is equivalent to the category of separable continuous
lattices and continuous maps. A similar theory yielding a category of
cpos was developed by Plotkin~\cite{PlotkinGD:tomud}. In this paper,
we will show that replacing the finite \emph{sets} $S$ in the above
construction with finite \emph{sequences} yields a category
appropriate for modelling imperative computation. 

The monoid in question has as its elements set of pairs $(s,n)$ where
$s$ is a finite sequence of natural numbers and $n$ is a
natural. Multiplication is defined by 
\[ a \cdot b = \{ (s_1\cdots s_k, n) \mid \exists m_1, \ldots,
m_k. (m_1\cdots m_k, n) \in b \land (s_i, m_i) \in a, i = 1, \ldots,
k\}
\]
where $s_1 \cdots s_k$ denotes the concatenation of the sequences
$s_1, \ldots, s_k$ and we identify singleton sequences with their
unique elements. 

Let us call this monoid $\mathcal{M}$ and its Karoubi envelope
$\mathcal{K}(\mathcal{M})$. Concretely, the connection between
$\mathcal{M}$ and Scott's monoid is very straightforward: sequences
replace Scott's finite sets, and concatenation replaces union.  It
seems obvious that the move from Scott's construction to ours is
nothing more than replacing one monad, the monad of finite powerset,
with another, that of finite sequences, in some formal
construction. In fact the situation is not quite so straightforward:
in order to set things up in an axiomatic fashion, one appears to
require a distributive law of the monad at hand over the powerset
monad. While the monad of finite sequences does distribute over
$\powerset$, $\powerset_\mathrm{fin}$ does not. This situation has
been studied by Hyland et al.\ in~\cite{PowerJ:cattfe}, where models
along the lines of Scott's are built axiomatically, using a
Kleisli-category construction. Their work only
applies to commutative monads, and therefore not to the
finite-sequence monad, so is not directly applicable here. Moreover,
for our purposes neither the category $\mathcal{K}(\mathcal{M})$ 
nor the kind of Kleisli construction proposed by Hyland et al. 
provides the most convenient setting in which to work. Although our
model of 
imperative computation can be seen as living entirely within these
categories, we shall propose a somewhat different
construction which yields additional structure useful in the analysis
of the model.

 We note also that Longley has recently shown how a
similar category, built from an untyped graph-style model using the
monad of finite multisets, as opposed to finite sets or finite
sequences, provides a model of fresh name
generation~\cite{LongleyJ:intlce}. In future work, we plan to
investigate the relationships between all these models in greater
detail, and explore the constructions at the higher level of
generality proposed by Hyland et al.

\section{Syntactic Control of Interference}
\label{sec:sci}
\noindent The imperative language we shall model is Reynolds's \emph{Syntactic
  Control of Interference} (\sci)~\cite{ReynoldsJC:synci}, and this
section is devoted to the presentation of its syntax, operational semantics
and notion of program equivalence. The
language was introduced by Reynolds as an approach to the problem of
establishing the non-interference properties of procedures and their
arguments required by specification logic. Reddy noticed that it was
precisely this interference-free fragment of an Algol-like language
which his model could interpret. Later, Reddy and O'Hearn showed that
the model could be extended to a full Algol-like language by means of
the Yoneda embedding~\cite{OHearnPW:objiye}, but it was not until the
refinement of game semantics was discovered that a fully abstract
model for such a language became available. 

The \sci\ language consists of a direct combination of the language of
while-loops, local variable allocation and the simply-typed
$\lambda$-calculus with an affine type discipline. The types of \sci\ 
are given by the grammar
\[
 A ::= \Nat \mid \Comm \mid \Var \mid A \llto A
\] 
where the base types are those of natural numbers ($\Nat$), commands
($\Comm$) and assignable variables ($\Var$). 
The terms of the language are as follows. 
\begin{eqnarray*}
 M & ::= & n \mid M + M \mid M - M \mid \ldots \\
   & \mid & \Skip \mid M \seq M \mid M \assign M \mid \deref M \\
   & \mid & \while{M}{M} \mid \ifzero{M}{M}{M} \\
   & \mid & x \mid \lambda x^A. M \mid MM \\
   & \mid & \new{x}{M} 
\end{eqnarray*}
where $n$ ranges over the natural numbers, $x$ over a countable set
of identifiers, and $A$ over the types of \sci. We adopt the usual
conventions with regard to binding of identifiers: $\lambda x^A.M$ binds $x$
in $M$; terms are identified up to $\alpha$-equivalence; and $M[N/x]$
denotes the capture-avoiding substitution of $N$ for free occurrences
of $x$ in $M$. 

The type system of the language imposes an affine discipline on
application: no function is allowed to share free identifiers with its
arguments. Typing judgments take the form
\[
 x_1:A_1, \ldots, x_n:A_n \vdash M:A
\]
where the $x_i$ are distinct identifiers, the $A_i$ and $A$ are
types, and $M$ is a term. We use $\Gamma$ and $\Delta$ to range over
contexts, that is, lists $x_1:A_1, \ldots, x_n:A_n$ of identifier-type
pairs with all identifiers distinct. The well-typed terms are given by
the following inductive definition, in which it is assumed that all
judgments are well-formed. 

%\textbf{TODO: weakening and exchange instead of this variable rule}
%DONE!!!

\begin{enumerate}[\hbox to8 pt{\hfill}]
\item\noindent{\hskip-12 pt\bf $\lambda$-calculus:}\
\[
\begin{prooftree}
\justifies
 x:A \vdash x:A
\end{prooftree}
\]

\[
\begin{prooftree}
  \Gamma, x:A \vdash M:B 
  \justifies
  \Gamma \vdash \lambda x^A. M:A \llto B
\end{prooftree}
\]

\[
\begin{prooftree}
  \Gamma \vdash M:A \llto B \quad \quad \Delta \vdash N:A 
\justifies
  \Gamma, \Delta \vdash MN:B
\end{prooftree}
\]
\item\noindent{\hskip-12 pt\bf Structural Rules:}\
\[
\begin{prooftree}
  \Gamma \vdash M 
\justifies
  \Gamma, x:A \vdash M
 \using{\mathrm{weakening}}
\end{prooftree}
\]

\[
\begin{prooftree}
 \Gamma \vdash M 
  \justifies
 \widetilde{\Gamma} \vdash M
  \using{\mathrm{exchange}}
\end{prooftree}
\]
\item\noindent{\hskip-12 pt\bf Arithmetic:}\
\[
\begin{prooftree}
\strut
 \justifies 
  \vdash n:\Nat
\end{prooftree}
\quad\quad
\begin{prooftree}
  \Gamma \vdash M:\Nat \quad\quad \Gamma \vdash N:\Nat 
  \justifies
  \Gamma \vdash M \odot N: \Nat 
 \using \odot \in \{ +, - , \ldots \}
\end{prooftree}
\]
\item\noindent{\hskip-12 pt\bf Sequential composition:}\
\[
\begin{prooftree}
\strut
  \justifies
  \vdash \Skip: \Comm
\end{prooftree}
\quad\quad
\begin{prooftree}
 \Gamma \vdash M: \Comm \quad \quad \Gamma \vdash N: B
 \justifies
 \Gamma \vdash M \seq N: B
  \using B \in \{\Comm, \Nat, \Var \}
\end{prooftree}
\]
\item\noindent{\hskip-12 pt\bf Assignable variables:}\
\[
\begin{prooftree}
  \Gamma \vdash M:\Var \quad \quad \Gamma \vdash N: \Nat 
\justifies
  \Gamma \vdash M \assign N: \Comm  
\end{prooftree}
\quad\quad
\begin{prooftree}
  \Gamma \vdash M: \Var 
\justifies
  \Gamma \vdash \deref M: \Nat
\end{prooftree}
\]
\item\noindent{\hskip-12 pt\bf Control structures:}\
\[
\begin{prooftree}
  \Gamma \vdash M: \Nat \quad \quad \Gamma \vdash N: \Comm 
 \justifies
  \Gamma \vdash \while{M}{N}: \Comm
\end{prooftree}
\]

\[
\begin{prooftree}
  \Gamma \vdash M: \Nat \quad \quad \Gamma \vdash N_1: B 
        \quad \quad \Gamma \vdash N_2: B 
 \justifies
  \Gamma \vdash \ifzero{M}{N_1}{N_2}: B
  \using B \in \{\Comm, \Nat, \Var \}
\end{prooftree}
\]
\item\noindent{\hskip-12 pt\bf Local blocks:}\
\[
\begin{prooftree}
  \Gamma, x:\Var \vdash M: B 
\justifies
  \Gamma \vdash \new{x}{M}: B 
    \using B \in \{\Comm, \Nat \}
\end{prooftree}
\]
\end{enumerate}

 In the exchange rule, $\widetilde{\Gamma}$ denotes any
permutation of the list $\Gamma$. 
In the rule for application, the assumption that the conclusion is
well-formed implies that $\Gamma$ and $\Delta$ contain distinct
identifiers. This was key to Reynolds's interference control agenda:
in the absence of a contraction rule, the only source of identifier
aliasing in the language is through procedure application, so by
enforcing the constraint that procedures and their arguments have no
identifiers in common, one eliminates all aliasing. It then follows
that program phrases with no common identifiers cannot interfere with
one another.

\paragraph{Note} Our version of \sci\ allows side-effects at all base types:
see the typing rule for sequential composition. We also include a
conditional at all base types. Variable allocation, however, is
restricted to blocks of type $\Comm$ and $\Nat$: terms such as
$\new{x}{x}$ are not permitted, because any sensible operational semantics
for such terms would violate the stack discipline for allocation and
deallocation of variables. 

The operational semantics of the language is given in terms of
\emph{stores}, that is, functions from identifiers to natural
numbers. A store $\sigma$ has as its domain a finite set of
identifiers, $\mathsf{dom}(\sigma)$. Given a store $\sigma$, we write
$\override{\sigma}{x}{n}$ for the 
store with domain $\mathsf{dom}(\sigma) \cup \{x\}$ which maps $x$
to $n$ and is identical to $\sigma$ on other identifiers. 
Note that this operation may extend the domain of $\sigma$. 

%\textbf{TODO: stores should be sigma not s}
% DONE!!!

Operational semantic judgments take the form
\[
 \Gamma \vdash \sigma, M \eval \sigma', V: A
\]
where
\begin{enumerate}[$\bullet$]
\item $\Gamma$ is a context containing only $\Var$-type identifiers
\item $\sigma$ and $\sigma'$ are stores whose domain is exactly those
  identifiers in $\Gamma$
\item $M$ and $V$ are terms
\item $A$ is a type
\item $\Gamma \vdash M: A$ and $\Gamma \vdash V:A$
\item $V$ is a \emph{value}, that is, a natural number, the constant
  $\Skip$, an identifier (which must have type $\Var$) or a
  $\lambda$-abstraction.  
\end{enumerate}
For the sake of brevity we omit the typing information from the
inductive definition below, writing judgments of the form 
$\sigma, M \eval \sigma', V$.  

\begin{enumerate}[\hbox to8 pt{\hfill}]
\item\noindent{\hskip-12 pt\bf Values and functions:}\
\[
\begin{prooftree}
\strut
\justifies
  \sigma, V \eval \sigma, V 
\using V~\mbox{a value}
\end{prooftree}
\quad\quad
\begin{prooftree}
\sigma, M\eval \sigma', \lambda x^A.M' 
\quad
\sigma', M'[N/x] \eval \sigma'', V
\justifies
\sigma, MN \eval \sigma'', V
\end{prooftree}
\]
\item\noindent{\hskip-12 pt\bf Operations:}\
\[
\begin{prooftree}
  \sigma, M_1 \eval \sigma', n_1 \quad \quad \sigma', M_2 \eval \sigma'', n_2 
 \justifies
  \sigma, M_1 \odot M_2 \eval \sigma'', n
 \using
 n = n_1 \odot n_2, \odot \in \{+, -, \ldots \}
\end{prooftree}
\]
\item\noindent{\hskip-12 pt\bf Variables:}\
\[
\begin{prooftree}
  \sigma, N \eval \sigma', n \quad \quad \sigma', M \eval \sigma'', x 
 \justifies
  \sigma, M \assign N \eval \override{\sigma''}{x}{n}, \Skip
\end{prooftree}
\quad\quad
\begin{prooftree}
  \sigma, M \eval \sigma', x
 \justifies
  \sigma, \deref M \eval \sigma', \sigma'(x)
\end{prooftree}
\]
\item\noindent{\hskip-12 pt\bf Control structures:}\
\[
\begin{prooftree}
  \sigma, M \eval \sigma', \Skip \quad \quad \sigma', N \eval \sigma'', V
 \justifies 
  \sigma, M \seq N \eval \sigma'', V
\end{prooftree}
\]

\[
\begin{prooftree}
  \sigma, M \eval \sigma', n 
\justifies
  \sigma, \while{M}{N} \eval \sigma',\Skip
 \using n \not = 0
\end{prooftree}
\]

\[
\begin{prooftree}
  \sigma, M \eval \sigma', 0 \quad \quad
  \sigma', N \eval \sigma'', \Skip \quad \quad
  \sigma'', \while{M}{N} \eval \sigma''', \Skip
\justifies
  \sigma, \while{M}{N} \eval \sigma''',\Skip
\end{prooftree}
\]

\[
\begin{prooftree}
  \sigma, M \eval \sigma', 0 \quad \quad
  \sigma', N_1 \eval \sigma'', V
\justifies
  \sigma, \ifzero{M}{N_1}{N_2} \eval \sigma'', V
\end{prooftree}
\]
\[
\begin{prooftree}
  \sigma, M \eval \sigma', n \quad \quad
  \sigma', N_2 \eval \sigma'', V
\justifies
  \sigma, \ifzero{M}{N_1}{N_2} \eval \sigma'', V
\using n \not = 0 
\end{prooftree}
\]
\item\noindent{\hskip-12 pt\bf Local blocks:}\
\[
\begin{prooftree}
  \override{\sigma}{x}{0}, M \eval \override{\sigma'}{x}{n}, V
 \justifies
  \sigma, \new{x}{M} \eval \sigma', V
\end{prooftree}
\]
\end{enumerate}

\noindent Note that in the rule for local blocks, the well-formedness
constraints on the conclusion $\sigma, \new{x}{M} \eval \sigma', V$
mean that the domains of definition of $\sigma$ and $\sigma'$ are the
same, and do not include $x$. Therefore the variable $x$ is only
available during the execution of the block $M$. 

We remark that, though the operational semantics takes account of the
possibility that evaluating a term of function-type could change the
store, the fact that all the store-changing term constructs are
confined to the base types means that this does not happen: whenever
$\sigma, M \eval \sigma', V$ for some $M$ and $V$ of type $A \llto B$,
we have $\sigma = \sigma'$ as a straightforward induction will
establish.

We now define a notion of \emph{contextual equivalence} on programs in
the usual way: given terms $\Gamma \vdash M, N: A$, we say that $M$
and $N$ are contextually equivalent, and write $M \cequiv N$, if and
only if for
every context $C[-]$ such that $\vdash C[M], C[N]: B$ for $B \in
\{\Comm, \Nat\}$, and every value $\vdash V:B$, 
\[
 C[M] \eval V \iff C[N]\eval V.
\]
(We omit the unique store over no variables from the operational
semantic judgments.)

One can also define a \emph{contextual preorder}: given the same data
as above, we write $M \capprox N$ iff for all contexts $C[-]$ and
values $V$, 
\[
 C[M] \eval V \Longrightarrow C[N]\eval V.
\]

\section{Reddy's object-spaces model}
\label{sec:object-spaces}
\noindent In this section we give a direct, concrete definition of a
semantics for \sci\ which accords with the model given by
Reddy~\cite{ReddyUS:gloscu}. To begin with we define the model without
imposing any structure on it, simply using sets and relations. Later
we go on to construct a category in which our modified graph model
lives as a monoid of endomorphisms of a particular object, and show
that the model of \sci\ inhabits that category. We shall then exploit
the structure of the category to obtain a clean proof of the model's
soundness. However, for pedagogical reasons we believe the concrete
presentation of the model in this section is worthwhile. In
particular, for the fragment of the language without abstraction and
application, the model is very simple and intuitively appealing, and
its soundness is easy to establish.

\subsection{A model based on events}

The key idea behind Reddy's model is that computations are interpreted
not as mappings from initial to final states (\ie \emph{state
  transformers}), but using sequences of observable \emph{events}. A
program will have as its denotation a set of tuples of such sequences. 

A type is interpreted as a set: the set of observable events at that
type. We define the semantics of types as follows.
\begin{eqnarray*}
  \sem{\Nat} & = & \naturals, & the set of natural numbers\\
  \sem{\Comm} & = & \{*\}, & a singleton set\\
  \sem{\Var} & = & \{\readmove(n), \writemove(n) \mid n \in
  \naturals\}\\
  \sem{A \llto B} & = & \sem{A}^* \times \sem{B}
\end{eqnarray*}
where $\sem{A}^*$ denotes the set of finite sequences over
$\sem{A}$.

The basic event one can observe of a term of type $\Nat$ is the
production of a natural number, so $\naturals$ is the interpretation
of $\Nat$. A closed term of type $\Comm$ can do nothing interesting
apart from terminating when executed, so $\Comm$ is interpreted as a
singleton set: we will see later that it is the open terms of type
$\Comm$ which behave more like state-transformers. At the type $\Var$,
there are two kinds of event: $\readmove(n)$ events correspond to
dereferencing a variable and receiving $n$ as the result, and
$\writemove(n)$ events correspond to assigning $n$ to the variable,
and observing termination of this operation. 

For the function types, the idea is that a single use of a function
$A\llto B$ will result in a single observable output event from $B$,
but may give rise to a sequence of events in the argument of type
$A$. Compare and contrast with Scott's $\pw$ model:
there functions are modelled as sets of pairs $(S,n)$ where $S$ is a
set of input-observations and $n$ is an output, while here we have
sets of pairs $(s,n)$ where the input observations form sequences rather than
sets. 

The denotation of a term 
\[
 x_1:A_1, \ldots, x_n:A_n \vdash M:B
\]
will be a set of tuples
\[
 (s_1, \ldots, s_n, b)
\]
where each $s_i \in \sem{A_i}^*$ and $b \in \sem{B}$. 
Again the idea is that such a tuple
records the ability of $M$ to produce observable event $b$ while
itself observing the sequences $s_i$ of events in (the terms bound to)
its free identifiers. 

\subsubsection{Remark}
Note that, in this model, the observed behaviour in each variable is
recorded separately; that is, there is no record of how interactions
with the various variables are interleaved. It is precisely this which
means we can only model \sci\ rather than the full Idealized Algol
language. The models based on game semantics refine the present model
by breaking each event into two, a start and a finish, and recording
the interleaving between actions, thereby overcoming this limitation. 

\vspace{2ex}

A little notation must be introduced before we give the definition of
the semantics. 
We will abbreviate such tuples $s_1, \ldots, s_n$ as $\vec{s}$, and 
semantic elements as above will become $(\vec{s}, b)$, or simply $b$
when $n=0$. We use $\vec{s}\vec{s'}$ 
to denote the componentwise concatenation of the tuples
of sequences $s_1, \ldots, s_n$ and $s'_1, \ldots, s'_n$. 

We say that a sequence $s \in \sem{\Var}^*$ is a \emph{cell-trace} iff
every $\readmove$ action in $s$ carries the same value as the most recent 
$\writemove$, if any, and zero if there has been no $\writemove$ yet.
(A formal definition appears later.)

%\textbf{TODO: weakening and contraction} 
% DONE!

We now give the definition of the semantics by induction on the typing
derivation of terms: for each typing rule, Figure~\ref{fig:semantics}
gives an equation which defines the semantics of the term in the
rule's conclusion by reference to the semantics of the terms in its
hypotheses.  

\begin{figure}
\begin{small}
\begin{eqnarray*}
%   \sem{x_1:A_1, \ldots, x_n:A_n \vdash x_i:A_i} & = & 
%   \{(\emptyseq_1,  \ldots, \emptyseq_{i-1}, a, \emptyseq_{i+1}, \ldots,
%   \emptyseq_n, a) \mid a \in \sem{A_i} \} \\
  \sem{x:A \vdash x:A} & = & \{ (a,a) \mid a \in \sem{A}\}\\
  \sem{\Gamma \vdash \lambda x^A. M:A \llto B} & = & 
\\
\makebox[0pt][l]{\hspace{-15ex}$
  \{
    (s_1, \ldots, s_n, (s,b)) \mid 
    (s_1, \ldots, s_n, s,b) \in \sem{\Gamma, x:A \vdash M:B}
  \}$}\\
  \sem{\Gamma, \Delta \vdash MN:B} & = & \\
\makebox[0pt][l]{\hspace{-15ex}$
  \left\{
    (\vec{s}, \vec{t^1}\ldots\vec{t^k},  b) \left|
    \begin{array}{l}
   \exists a_1, \ldots, a_k. 
(\vec{s}, (a_1 \ldots a_k, b)) \in \sem{\Gamma \vdash
     M:A\llto B}\\
   \land (\vec{t^i}, a_i) \in \sem{\Delta\vdash N:A}~\mbox{for}~i=1,\ldots,k \\ 
  \end{array}\right.
\right\}$} \\
  \sem{\Gamma, x:A\vdash M:B} & = & \{(\vec{s}, \emptyseq, b) \mid
  (\vec{s},b)\in\sem{\Gamma \vdash M:B} \}\\
  \sem{\widetilde{\Gamma} \vdash M:A} & =  & \{ (\widetilde{\vec{s}},
  a) \mid (\vec{s},a) \in \sem{\Gamma \vdash M:A} \}\\
  \sem{\vdash n:\Nat} & = & \{n\}\\
  \sem{\Gamma \vdash M_1 \odot M_2:\Nat} & = & \\
\makebox[0pt][c]{\hspace{35ex}$
 \{ (\vec{s} \vec{s'}, m_1 \odot m_2) \mid 
    (\vec{s}, m_1) \in \sem{\Gamma\vdash M_1:\Nat}, (\vec{s'}, m_2)\in
    \sem{\Gamma \vdash M_2:\Nat}\}$} \\
  \sem{\vdash \Skip:\Comm} & = & \{*\}\\
\sem{\Gamma \vdash M \seq N: B} & = & \\
\makebox[0pt][l]{\hspace{-15ex}$
 \{ (\vec{s} \vec{s'}, b) \mid 
    (\vec{s}, *) \in \sem{\Gamma \vdash M:\Comm}, (\vec{s'}, b) \in
    \sem{\Gamma \vdash N:B}\}$} \\
  \sem{\Gamma \vdash M \assign N} & = & \\
\makebox[0pt][l]{\hspace{-15ex}$
 \{ (\vec{s} \vec{s'}, *) \mid 
    (\vec{s}, n) \in \sem{\Gamma \vdash N:\Nat}, (\vec{s'},
    \writemove(n)) \in \sem{\Gamma\vdash M:\Var}\}$} \\
  \sem{\Gamma \vdash \deref M:\Nat} & = & 
 \{ (\vec{s}, n) \mid 
    (\vec{s}, \readmove(n)) \in \sem{\Gamma \vdash M:\Var}\} \\
  \sem{\Gamma \vdash \while{M}{N}:\Comm} & = & \\
\makebox[0pt][l]{\hspace{-15ex}$
 \left\{
    (\vec{s^1}\vec{t^1}\vec{s^2}\vec{t^2}\ldots\vec{s^j}\vec{t^j}\vec{s}, *) 
     \left|
 \begin{array}{l}
    \forall i. (\vec{s^i}, 0) \in \sem{\Gamma \vdash M:\Nat}\\
 \land  (\vec{t^i}, *) \in \sem{\Gamma \vdash N:\Comm} \\ 
 \land \exists m \not = 0.  (\vec{s}, m) \in \sem{\Gamma \vdash M:\Nat}
 \end{array}\right.
 \right\}$} \\
\sem{\Gamma \vdash \ifzero{M}{N_1}{N_2}:B}  & = & \\
\makebox[0pt][l]{\hspace{-15ex}$
   \{ 
    (\vec{s}\vec{t}, b) \mid
    (\vec{s}, 0) \in \sem{\Gamma \vdash M:\Nat}, 
    (\vec{t}, b) \in \sem{\Gamma \vdash N_1:B}
   \}$} \\ 
  &  \cup & \\
\makebox[0pt][l]{\hspace{-15ex}$
   \{ 
    (\vec{s}\vec{t}, b) \mid \exists m \not = 0. 
    (\vec{s}, m) \in \sem{\Gamma \vdash M:\Nat}, 
    (\vec{t}, b) \in \sem{\Gamma \vdash N_2:B}
   \}$} \\
  \sem{\Gamma \vdash \new{x}{M}:B} & = & 
  \left\{
     (\vec{s}, b) \left|
  \begin{array}{l}
     \exists s. (\vec{s}, s, b) \in \sem{\Gamma, x:\Var \vdash M:B}
 \\   \land~s~\mbox{is a cell trace}. 
   \end{array}\right.
       \right\}
\end{eqnarray*}
\end{small}
\caption{Reddy-style semantics of \sci}
  \label{fig:semantics}
\end{figure}

\subsection{Examples}
\begin{enumerate}[$\bullet$]
\item Consider the program $\mathsf{swap}$, defined by 
\[
 x: \Var, y:\Var, z:\Var \vdash z \assign \deref x \seq x \assign \deref
 y \seq y \assign \deref z: \Comm.
\]
It is straightforward to compute that $\sem{\mathsf{swap}}$ is the set
\[
 \{ (\readmove(n)\writemove(n'), \readmove(n')\writemove(n''),
 \writemove(n)\readmove(n''), *) \mid n, n', n'' \in \naturals
 \}. 
\]
The semantic definitions do not yet enforce variable-like behaviour,
so that in particular $n$ and $n''$ need not be equal. 

However, the semantics of $\new{z}{\mathsf{swap}}$ selects just those
entries in which $z$ behaves like a good variable, so that $n = n''$,
and then hides the $z$-behaviour: 
\[
 \sem{\new{z}{\mathsf{swap}}} = 
 \{ (\readmove(n)\writemove(n'), \readmove(n')\writemove(n), *)
    \mid n, n' \in \naturals
 \}. 
\]
Thus the values in $x$ and $y$ are swapped, and the semantics does
not record anything about the use of $z$ or the fact that $x$ was
reassigned first. 

\item The type $\Comm \llto \Comm$ has as its elements all pairs of the
  form
\[
 (  *\cdot *\cdot *\cdots *, *). 
\]
  A deterministic program of this type will contain at most one such
  element in its denotation, corresponding to a ``for loop'' which
  executes its argument a fixed, finite number of times. There is also
  the empty set, corresponding to a program which never terminates
  regardless of its argument. 
\end{enumerate}

\subsection{Soundness for the ground types}
\label{sec:ground-sound}
We now prove that our model is sound with
respect to the operational semantics for the fragment of the language
excluding abstraction, application, and non-base types. We refer to
this fragment as \scim; it is essentially the language of
while-programs plus block allocated variables. 

First let us introduce a little more notation. 

We define a notion of state transition. Given a sequence $s \in
\sem{\Var}^*$, we define the transitions
\[
 \strans{n}{s}{n'}
\]
where $n$ and $n'$ are natural numbers, as follows. 
\[
\begin{prooftree}
 \justifies
 \strans{n}{[]}{n}
\end{prooftree} 
\quad \quad
\begin{prooftree}
 \justifies
 \strans{n}{[\readmove(n)]}{n}
\end{prooftree} 
\]
\[
\begin{prooftree}
 \justifies
 \strans{n}{[\writemove(n')]}{n'}
\end{prooftree} 
\quad \quad
\begin{prooftree}
  \strans{n}{s}{n'} \quad \strans{n'}{s'}{n''}
 \justifies
 \strans{n}{ss'}{n''}
\end{prooftree} 
\]
We write $\strans{n}{s}{}$ to mean that $\strans{n}{s}{n'}$ for some
$n'$. We can now give a precise definition of cell-trace: a sequence
$s \in \sem{\Var}^*$ is a cell-trace if and only if
$\strans{0}{s}{}$. Note also that $\strans{n}{s}{}$ if and only if
$\writemove(n)s$ is a cell-trace. 

We extend this to traces involving more than one $\Var$ type as
follows. Given a context $x_1: \Var, \ldots, x_n: \Var$, an element
$s = (s_1, \ldots, s_n) \in \sem{\Var}^* \times \cdots \times \sem{\Var}^*$,
and stores $\sigma$ and $\sigma'$ in variables $x_1$, \ldots, $x_n$,
we write
\[
 \strans{\sigma}{s}{\sigma'}
\]
iff 
\[
  \strans{\sigma(x_i)}{s_i}{\sigma'(x_i)}
\]
for each $i$.

\begin{definition}
  Say that a term $\Gamma \vdash M: B$, where $B$ is a base type and
  $\Gamma$ contains only $\Var$-typed variables, is \emph{good} if and
  only if:
\begin{enumerate}[\hbox to8 pt{\hfill}]
\item\noindent{\hskip-12 pt\bf Case $B=\Comm$:}\
  for all stores $\sigma, \sigma'$ over $\Gamma$, 
\[
   \sigma, M \eval \sigma', \Skip \Leftrightarrow
   \exists (\vec{s},*) \in
   \sem{M}. \strans{\sigma}{\vec{s}}{\sigma'}
\]
  \item\noindent{\hskip-12 pt\bf Case $B=\Nat$:}\
   for all stores $\sigma, \sigma'$ over $\Gamma$
    and all $n \in \naturals$, 
\[
   \sigma, M \eval \sigma', n \Leftrightarrow
   \exists (\vec{s},n) \in
   \sem{M}. \strans{\sigma}{\vec{s}}{\sigma'}
\]
  \item\noindent{\hskip-12 pt\bf Case $B=\Var$:}\
   $\Gamma \vdash \deref M: \Nat$ is good and 
   for all $n \in \naturals$, $\Gamma \vdash M \assign n: \Comm$ is
   good. 
  \end{enumerate}
\end{definition}

\begin{lem}
\label{lem:scim-good}
  All terms $\Gamma \vdash M:B$ of \scim, where $B$ is a base type and $\Gamma$
  contains only $\Var$-typed variables, are good in the above sense. 
\end{lem}
\proof
  We proceed by induction on the structure of the term $M$. For the
  constants $\Skip$ and $n$, the result is trivial. For variables
  $x:\Var$, we must show that both $\deref x$ and $x \assign n$ are
  good. 

  Unpacking the definitions, we have
\[
  \sem{\deref x} = 
  \{(\vec{\emptyseq}, \readmove(n), \vec{\emptyseq}, n) \mid n \in
  \naturals \}. 
\]
  But 
  $\strans{\sigma}{\vec{\emptyseq}, \readmove(n),
    \vec{\emptyseq}}{\sigma'}$
   if and only if $\sigma = \sigma'$ and $\sigma(x) = n$, which holds
   if and only if $\sigma, \deref x \eval \sigma', n$. 

  For the assignment part, we have
\[
  \sem{x \assign n} = 
  \{(\vec{\emptyseq}, \writemove(n), \vec{\emptyseq}, *) \}
\]
  and 
  $\strans{\sigma}{\vec{\emptyseq}, \writemove(n),
    \vec{\emptyseq}}{\sigma'}$
   if and only if $\sigma' = \override{\sigma}{x}{n}$, which holds
   if and only if $\sigma, x \assign n \eval \sigma', \Skip$. 

  For $\while{M}{N}$, first note that 
\[
 \sigma, \while{M}{N} \eval \sigma', \Skip
\]
  if and only if there are sequences of stores $\sigma_i$ and
  $\tau_i$, for $i = 1, \ldots, n$, such that $\sigma = \sigma_1$,
  $\sigma' = \tau_n$, 
  \[
   \sigma_i, M \eval \tau_i, 0 \quad \quad \tau_i, N \eval
   \sigma_{i+1}, \Skip
  \]
  for $i = 1, \ldots, n-1$ and 
  \[
    \sigma_n, M \eval \tau_n, k 
  \]
  for some $k \not = 0$. 
  (This can be proved by induction on derivations in the
  operational semantics of $\mathtt{while}$.)

  Therefore, applying the inductive hypothesis to $M$ and $N$, we have
  that 
\[
 \sigma, \while{M}{N} \eval \sigma', \Skip
\]
  if and only if there are $\vec{s_1}, \ldots, \vec{s_n}$ and $\vec{t_1},
    \ldots, \vec{t_{n-1}}$ such that 
  \[
   (\vec{s_i}, 0) \in \sem{M} \quad \quad (\vec{t_i}, *) \in \sem{N}
  \]
   for $i = 1, \ldots, n-1$ and 
  \[
    (\vec{s_n}, k) \in \sem{M}
  \]
  for some $k\not = 0$, and moreover 
  \[
   \strans{\sigma_i}{\vec{s_i}}{\tau_i} \quad \quad
   \strans{\tau_i}{\vec{t_i}}{\sigma_{i+1}}
  \]
  for $i = 1, \ldots, n-1$ and 
  \[
   \strans{\sigma_n}{\vec{s_i}}{\tau_n}. 
  \]
  But then we have that
  \[
   \strans{\sigma_1}{\vec{s_1}\vec{t_1}\ldots\vec{s_{n-1}}\vec{t_{n-1}}\vec{s_n}}{\tau_n} 
  \]
  and 
  \[
  (\vec{s_1}\vec{t_1}\ldots\vec{s_{n-1}}\vec{t_{n-1}}\vec{s_n}, *) \in
  \sem{\while{M}{N}}
  \]
  by definition. Furthermore, all elements of $\sem{\while{M}{N}}$
  with cell-traces in the $\Gamma$ part are of this form, which
  establishes the converse.  

  The case of $\ifzero{M}{N_1}{N_2}$ is similar to this one, and
  simpler. 

  Consider the case of $M \assign N$. By definition of the operational
  semantics, 
  \[
  \sigma, M \assign N \eval \sigma', \Skip
  \]
  if and only if there are $\sigma''$, $\sigma'''$, $x$ and $n$ such that
  \[
   \sigma, N \eval \sigma'', n \quad\quad \sigma'', M \eval \sigma''',
   x
  \]
  and $\sigma' = \override{\sigma'''}{x}{n}$. This is the same as
  saying 
\begin{equation}
   \sigma, N \eval \sigma'', n \quad\quad \sigma'', M \assign n \eval
   \sigma', \Skip. 
\label{eq:assigns}
\end{equation}
By the inductive hypothesis, both $N$ and $M$ are good, and hence by
  definition of ``good'' for terms of type $\Var$, $M \assign n$ is
  good, so~(\ref{eq:assigns}) holds if and only if we have
  \begin{equation}
  (\vec{s}, n) \in \sem{N} \quad \quad (\vec{t},*) \in \sem{M \assign
  n} \label{eq:assignment}. 
  \end{equation}
  such that 
  \[
   \strans{\sigma}{\vec{s}}{\sigma''} \quad \quad
   \strans{\sigma''}{\vec{t}}{\sigma'}.
  \]
  By definition of the semantics, 
  \[
 (\vec{t},*) \in \sem{M \assign n} \Leftrightarrow
  (\vec{t},\writemove(n)) \in \sem{M}
  \]
  so~(\ref{eq:assignment}) holds if and only if 
  \[
   (\vec{s} \vec{t}, *) \in \sem{M \assign N}. 
  \]
  
  The case of $\deref M$ follows directly from the inductive
  hypothesis: since $M$ is good, so is $\deref M$. 

  Finally we consider $\new{x}{M}: \Comm$ (the $\Nat$ case is
  similar). 
  By definition of the operational semantics, 
  \[
  \sigma, \new{x}{M} \eval \sigma', \Skip
  \]
  iff
  \[
   \override{\sigma}{x}{0}, M \eval \override{\sigma'}{x}{n}, \Skip. 
  \]
  By the inductive hypothesis, this is possible if and
  only if there is some $(\vec{s}, s', *) \in \sem{M}$ with 
  \[
   \strans{\sigma}{\vec{s}}{\sigma'} \quad \quad 
   \strans{0}{s'}{n}. 
  \]
  The second condition above is the definition of $s'$ being a cell-trace, so
  this holds if and only if $(\vec{s},*) \in \sem{\new{x}{M}}$ as
    required. 
\qed

The fact that all terms are good gives us the following soundness
result for \scim. 

\begin{cor}
\label{cor:scim-sound}
For any closed term $\vdash M: B$ of \scim, where $B$ is $\Comm$ or $\Nat$, 
$M\eval V$ if and only if $\sem{M} = \sem{V}$. \qed
\end{cor}

\section{A category of monoids and relations}

\noindent Before going on to establish the soundness of Reddy's model for the
whole of \sci, we shall develop a categorical setting for the model,
based on monoids and relations. Our monoid $\mathcal{M}$ 
appears as the monoid of endomorphisms of an object in this category,
so the retracts of this object all live in the category
$\mathcal{K}(\mathcal{M})$. It happens that all the objects we use to
interpret types of $\sci$ are indeed retracts of this object, so the
graph construction does indeed yield a category suitable for modelling
imperative computation. Nevertheless it is useful to describe the
larger category. Not only is its construction straightforward, but
also it possesses some structure beyond that of
$\mathcal{K}(\mathcal{M})$ which makes the description of Reddy's
model more straightforward, and allows the soundness result above to
be extended to the whole language using algebraic reasoning. 

We believe that there is a more general description of these
constructions to be found, perhaps extending the work
of~\cite{PowerJ:cattfe}; but we leave this for future work. 

To build our category, we will be making use of the category $\mon$ of
monoids and homomorphisms, and exploiting the product, coproduct and
powerset operations on monoids, and the notion of the free monoid over
a set. For the sake of completeness, we review these constructions
here.

First some notation. For a monoid $A$, we use
$e_A$ to denote the identity element, and write monoid multiplication
as concatenation, or occasionally using the symbol $\cdot_A$. The
underlying set of the monoid $A$ is written as $UA$. 

\subsubsection{Free monoids}
Recall that for any set $A$, the \emph{free monoid over $A$} is given
by $A^*$, the monoid of strings over $A$, also known as the Kleene
monoid over $A$. The operation taking $A$ to $A^*$ is left-adjoint to the
forgetful functor $U: \mon \rightarrow \mathbf{Set}$. 

\subsubsection{Products}
The category $\mon$ has products. The product of monoids $A$ and $B$
is a monoid with underlying set $UA \times UB$, the Cartesian product
of sets. The monoid operation is defined by 
\[
 \pair{a}{b}\pair{a'}{b'} = \pair{a \cdot_A a'}{b \cdot_B b'}. 
\]
The identity element is $\pair{e_A}{e_B}$. Projection and pairing maps
in $\mon$ are given by the corresponding maps on the underlying
sets. The terminal object is the one-element monoid. The construction
given above generalizes to give all small products. 

\subsubsection{Coproducts}
The category $\mon$ also has finite coproducts. These are slightly awkward
to define in general, and since we will not be making use of the
general construction, we omit it here. 

The special case of the coproduct of two free monoids is easy to
define. Since the operation of building a free monoid from a set is
left adjoint to the forgetful functor $U$,  it preserves
colimits and in particular coproducts. For sets $A$ and $B$,
the coproduct monoid $A^* + B^*$ is therefore given by $(A+B)^*$, the
monoid of strings over the disjoint union of $A$ and $B$. 

The initial object is the one-element monoid. 

\subsubsection{Powerset}
The familiar powerset construction on $\mathbf{Set}$ lifts to $\mon$
and retains much of its structure. Given a monoid $A$, define the
monoid $\powerset A$ as follows. Its underlying set is the powerset of
$UA$, that is, the set of subsets of $UA$. Monoid multiplication is
defined by
\[
 ST = \{ x\cdot_A y \mid x \in S, y \in T \}
\]
and the identity is the singleton set $\{e_A\}$. 

We will make use of
the Kleisli category $\mon_\powerset$. This category can be defined
concretely as follows. Its objects are monoids, and a map from $A$ to
$B$ is a monoid homomorphism from $A$ to $\powerset B$. The identity
on $A$ is the singleton map which takes each $a \in A$ to
$\{a\}$. Morphisms are composed as follows: given maps $f:A
\rightarrow B$ and $g: B \rightarrow C$, the composite $f\comp g:A
\rightarrow C$ is defined by 
\[
 (f; g)(a) = \{ c \mid \exists b \in f(a). c \in g(b) \}.
\]

The fact that powerset is a \emph{commutative monad}
on $\mon$  means that the product 
structure on $\mon$ lifts to a monoidal structure on
$\mon_\powerset$ as follows. We define $A \tensor B$ to be
the monoid  $A \times B$. For the functorial action, we make use of
the \emph{double strength} map
\[
 \theta_{A,B}: \powerset A \times \powerset B \longrightarrow
 \powerset(A\times B)
\]
defined by 
\[
 \theta_{A,B}(S,T) = \{ \pair{x}{y}\mid x \in S, y \in T \}. 
\]
This is a homomorphism of monoids. With this in place, given maps $f:A
\rightarrow B$ and $g:C \rightarrow D$ in $\mon_\powerset$, we can
define $f \tensor g : A \tensor C \rightarrow B \tensor D$ as the
homomorphism $f \times g \comp \theta_{B,D}$. See for
example~\cite{JacobsB:semwc} for more details on this construction. 

\subsection{The category}
The category we will use to model \sci\ is
$\op{(\mon_\powerset)}$. This category can be seen as a 
category of ``monoids and relations'' of a certain kind, so we will
call it $\monrel$. 

We now briefly explore some of the structure that $\monrel$
possesses. 

\subsubsection{Monoidal structure}
The monoidal structure on $\mon_\powerset$ described above
is directly inherited by $\monrel$.  Furthermore,
since the unit $I$ of the monoidal structure is given by the
one-element monoid, which is also an initial object in $\mon$, $I$ is
in fact a terminal object in $\monrel$, so the category has an
\emph{affine} structure. An important consequence of this is that
projections exist: for any $A_1$, \ldots, $A_n$ there are canonical maps
\[
 \pi_i: A_1 \tensor \cdots \tensor A_n \rightarrow A_i.
\]

\subsubsection{Exponentials}
Let $A$ and $B$ be any monoids, and $C^*$ be the free monoid over some
set $C$. Consider the following sequence of natural isomorphisms and
definitional equalities.
\begin{eqnarray*}
  & & \monrel(A \tensor B, C^*) \\
 & = & \mon(C^*, \powerset(A \times B)) \\
& \cong & \mathbf{Set}(C, U\powerset(A \times B)) \\
& \cong & \rel(C, UA \times UB) \\
& \cong & \rel(UB \times C, UA)
\end{eqnarray*}
Similarly we can show that 
\[
 \rel(UB \times C, UA) \cong \monrel(A, (UB \times C)^*).
\]
The exponential $B \llto C^*$ is therefore given by 
$(UB \times C)^*$. It is important to note that the free monoids are
closed under this operation, so that we can form 
$A_1 \llto (A_2 \llto \ldots (A_n \llto C^*))$ for any $A_1$, \ldots,
$A_n$. That is to say, the free monoids form an \emph{exponential
  ideal} in $\monrel$. 

Given a map $f: A \tensor B \rightarrow C^*$ in $\monrel$, we write
$\Lambda(f)$ for the curried map $A \rightarrow (B \llto C^*)$. 
The counit of the adjunction is written
\[
\ev: (B \llto C^*) \tensor B \rightarrow C^*. 
\]

\subsubsection{Products}
The coproduct in $\mon$ is inherited by the Kleisli-category
$\mon_\powerset$, and since $\monrel$ is the opposite of this
category, $\monrel$ has products. 

\subsubsection{An alternative characterization}
We can also describe the category $\monrel$ concretely, as follows. 
Objects are monoids, and maps $A \rightarrow B$ are relations $R$
between (the underlying sets of) $A$ and $B$, with the following
properties:  
\begin{enumerate}[\hbox to8 pt{\hfill}]
\item\noindent{\hskip-12 pt\bf  homomorphism:}\
   $e_A R e_B$, and if $a_1 R b_1$ and $a_2 R b_2$, then
   $a_1 a_2 R b_1 b_2$
\item\noindent{\hskip-12 pt\bf identity reflection:}\
   if $a R e_B$ then $a = e_A$
\item\noindent{\hskip-12 pt\bf decomposition]:}\
  if $a R b_1 b_2$ then there exist $a_1, a_2 \in
  A$ such that $a_i R b_i$ for $i = 1,2$ and $a = a_1 
  a_2$. 
\end{enumerate}
  Identities and composition are as usual for relations. 
  Note that the property of ``identity reflection'' is merely the
  nullary case of the property of ``decomposition''.

It is routine to show that this definition yields a category
isomorphic to $\op{(\mon_\powerset)}$. The action of the isomorphism is
as follows. Given a map $A \rightarrow B$ in $\op{(\mon_\powerset)}$,
that is to say, a homomorphism 
\[
 f: B \longrightarrow \powerset(A)
\]
we can define a relation $R_f$ between $A$ and $B$ as the set of pairs
$\{ (a,b) \mid a \in f(b)\}$. 

\subsubsection{Recovering the monoid $\mathcal{M}$}
We remark that the monoid of endomorphisms of the object $\omega^*$,
the monoid of sequences of natural numbers, is exactly the monoid
$\mathcal{M}$ of Section~\ref{sec:scott-pw}. A map $\omega^*
\rightarrow \omega^*$ consists of a monoid homomorphism $\omega^*
\rightarrow \powerset\omega^*$ which is the same as an ordinary
function $\omega \rightarrow \powerset\omega^*$. Reversing the arrows
and using the characterization of $\rel$ as the Kleisli-category for
$\powerset$ on $\mathbf{Set}$, this is just a subset of $\omega^* \times
\omega$, and it is routine to check that the composition of these sets
is as described in Section~\ref{sec:scott-pw}. 

It follows that the full subcategory of $\monrel$ consisting of
objects which are retracts of
$\omega^*$ can also be seen a subcategory of the Karoubi envelope
$\mathcal{K}(\mathcal{M})$, and it will turn out that all the types of
\sci\  are modelled using objects of this subcategory. Just as Scott
used the Karoubi envelope of $\pw$ as a category for giving semantics,
we can use $\mathcal{K}(\mathcal{M})$. However, $\monrel$ proves to be
a more convenient category, because it possesses additional objects,
in particular tensor products such as $\omega^* \lltensor \omega^*$,
which assist in the description and analysis of our model but do not
belong to $\mathcal{K}(\mathcal{M})$. 

It is perhaps worth remarking that Reddy's original work struggled to
find a satisfying categorical setting for the model, resorting to the
use of multicategories in the absence of objects such as $\omega^*
\lltensor \omega^*$. We believe our new categorical setting paints a
more convincing picture. 

\subsection{Modelling \texorpdfstring{\sci}{SCI}  in \texorpdfstring{$\monrel$}{MonRel}}

We now show how Reddy's model of \sci\ lives in $\monrel$. Types are
interpreted as objects of the category, that is, as monoids. Indeed
every type is interpreted as the free monoid over the set which we
used for the direct presentation of the semantics given
above. Formally we can give an inductive definition of the semantics
of types as follows. 
\begin{eqnarray*}
  \sem{\Comm} & = & 1^*\\
  \sem{\Nat} & = & \naturals^*\\
  \sem{\Var} & = & \sem{\Comm}^\omega \times \sem{\Nat} \\
  \sem{A \llto B} & = & \sem{A} \llto \sem{B}. 
\end{eqnarray*}
For the definition of $\sem{A \llto B}$ to make sense it is essential
that every $\sem{B}$ is a free monoid. This is clear for the base
types $\Comm$ and $\Nat$.  Recalling that products in $\monrel$ come
from coproducts in $\mon$, and that the coproduct of free monoids is
again a free monoid, we see that $\sem{\Var}$ is a free monoid, and
therefore by induction every types is interpreted as the free monoid over
some alphabet. 

Let us write $\alpha A$ for the underlying alphabet of $\sem{A}$, and
verify that for every type $A$, $\alpha A$ is the set that was used in
the direct presentation of the semantics above. 

For $\Comm$ and $\Nat$, this is clear. 
To see that the same holds for $\Var$, recall that products in
$\monrel$ come from coproducts in $\mon$, which for free monoids are
given by disjoint union of alphabets. 
So 
\[
 \alpha \Var = \left(\sum_w 1 \right)+ \naturals. 
\]
The single element of the $n$th summand of the left component
corresponds to $\writemove(n)$, and the element $n$ of the right
component corresponds to $\readmove(n)$; indeed we will continue to
use this notation below. Our reason for giving the 
semantic definition in the above form will become clear when we come
to the semantics of assignment and dereferencing. 

Finally, by the definition of exponential, 
\[
 \alpha (A \llto B) = (\alpha A)^* \times \alpha B
\]
which agrees with our previous definition. 

For the semantics of terms, we exploit the categorical structure of
$\monrel$: the $\lambda$-calculus part is interpreted using the
monoidal and exponential structure of the category, while the
constants are interpreted by defining particular maps in the category,
making use of products for those constants which allow their operands
to share variables. 

A term $x_1: A_1, \ldots, x_n:A_n \vdash M: B$ is interpreted as a map
\[
 \sem{M}: \sem{A_1} \tensor \cdots \tensor \sem{A_n} \rightarrow
 \sem{B}. 
\]
(If $\Gamma$ is the context $x_1:A_1, \ldots, x_n:A_n$ we will often
abbreviate the object $\sem{A_1} \tensor \cdots \tensor \sem{A_n}$ as
$\sem{\Gamma}$). 
Unpacking definitions, such a map is a homomorphism
\[
 \sem{B} \rightarrow \powerset(\sem{A_1} \times \cdots \times
 \sem{A_n}). 
\]
Since all types are interpreted as free monoids, 
this is the same as an ordinary function
\[
 \alpha B \rightarrow \powerset((\alpha A_1)^* \times \cdots \times
 (\alpha A_n)^* )
\]
which in turn corresponds to a subset of 
\[
 (\alpha A_1)^* \times \cdots \times (\alpha A_n)^* \times \alpha B. 
\]
Under this representation, the denotations of terms in $\monrel$ have
the same form as those in the direct presentation, and we will use the
``sets of tuples'' when we need to define morphisms explicitly. 

% \textbf{TODO: weakening and exchange}
% DONE!!!

A variable is interpreted as the identity map:
\[
 \sem{x:A \vdash x:A} = \mathsf{id}:\sem{A} \rightarrow \sem{A}. 
\]
Weakening is interepreted using projections: if 
\[\sem{\Gamma \vdash M:B} = f:\sem{\Gamma} \rightarrow \sem{B}\]
then 
\[
\sem{\Gamma, x:A \vdash M:B} = \pi \seq f
\]
where $\pi: \sem{\Gamma} \lltensor \sem{A} \rightarrow \sem{\Gamma}$
is a projection map. 

Exchange is interpreted using the symmetry isomorphisms: for any
permutation on a context taking $\Gamma$ to $\widetilde{\Gamma}$ there
is a corresponding isomorphism $\mathsf{symm}: \sem{\widetilde{\Gamma}}
\rightarrow \sem{\Gamma}$, and then 
\[
 \sem{\widetilde{\Gamma}\vdash M:A} = \mathsf{symm} \seq \sem{\Gamma \vdash
   M:A}. 
\]

Abstraction is interpreted using the currying part of the exponential
adjunction: if 
\[
  \sem{\Gamma, x:A \vdash M:B} = f: \sem{\Gamma} \tensor \sem{A} \rightarrow \sem{B}
\]
then
 \[
  \sem{\Gamma \vdash \lambda x^A. M : A \llto B} = \Lambda(f): \sem{\Gamma}\rightarrow
  \sem{A} \llto \sem{B}.  
 \]

Application is interpreted using $\ev$: 
\[
 \sem{MN} = \sem{M} \tensor \sem{N} \comp \ev. 
\]

It is straightforward to check that these definitions agree with the
concrete ones given earlier. 

To interpret the basic imperative constructs, we define a collection of
maps in the category. For instance, to interpret $\while{M}{N}$ we use a map
\[
   w: \sem{\Nat} \times \sem{\Comm} \rightarrow \sem{\Comm}
\]
which we will define below, and set
\[
 \sem{\while{M}{N}} = \pair{\sem{M}}{\sem{N}} \comp w. 
\]
The object $\sem{\Nat} \times \sem{\Comm}$ is the free monoid over the
alphabet $\naturals \cup \{*\}$. We can therefore define $w$ as the
set of tuples
\[
 w = \{ (0 * 0 * \cdots 0 * n, *)  \mid n \not = 0 \}. 
\]
Maps interpreting $\ifzero{M}{N_1}{N_2}$, $\deref M$ and $M \assign N$
can be defined similarly and all yield interpretations which agree
with the direct one. However, for assignment and dereferencing, the
definition of $\sem{\Var}$ as $\sem{\Comm}^\omega \times \sem{\Nat}$
suggests a more abstract definition using projections: there are
projections
\[
 \mathsf{assign}(n): \sem{\Var} \rightarrow \sem{\Comm}
\]
for each $n$, and
\[
 \mathsf{deref}: \sem{\Var} \rightarrow \sem{\Nat}
\]
and these are indeed the maps we need. Thus our interpretation of
$\Var$ has the kind of ``object oriented'' flavour advocated by
Reynolds: a variable is an object with $\omega$-many write-methods and
a read-method, and its semantics is given by the product of these. 

Finally the semantics of $\mathtt{new}$ is given by means of maps of type
\[
 \sem{\Var \llto \Comm} \rightarrow \sem{\Comm}
\quad\quad
\mbox{and}
\quad\quad
 \sem{\Var \llto \Nat} \rightarrow \sem{\Nat}
\]
defined by the sets
\[
 \{ ((s,*), *) \mid s~\mbox{is a cell trace}\}
\]
and
\[
 \{ ((s,n), n) \mid n \in \naturals, s~\mbox{is a cell trace}\}
\]
respectively.

\subsection{Soundness of the model of \texorpdfstring{\sci}{SCI}}

We can now show that our model is sound for the whole of \sci,
extending the result of Section~\ref{sec:ground-sound}. 

First a standard lemma which says that substitution is modelled by
composition in the category.
\begin{lem}[Substitution]
\label{lem:sub}
  If $\Gamma, x:A \vdash M:B$ and $\Delta \vdash N: A$ are terms of
  \sci, then so is $\Gamma, \Delta \vdash M[N/x]: B$, and furthermore
$ \sem{M[N/x]} = \id_{\sem{\Gamma}} \tensor \sem{N} ; \sem{M}.$\qed
\end{lem}

With this in place it is standard that $\beta$-reduction is soundly
modelled, because of the naturality of currying.
\begin{lem}
\label{lem:beta}
  If $\Gamma, x:A \vdash M:B$ and  $\Delta \vdash N: A$, then
$ \sem{(\lambda x. M)N} = \sem{M[N/x]}. $\qed
\end{lem}

Both of these Lemmas are proved by a straightforward induction on the
structure of terms. They hold for standard reasons, because we are
working in a symmetric monoidal category and using exponentials to
model function spaces. We can now establish soundness for our model
using purely algebraic reasoning: the fact that there is no recursion
in the language makes this particularly straightforward. The key is to
establish that every ground-type term of the full language has the
same behaviour as a term of $\scim$; a property that is captured by
the following definition. 

\begin{definition}
  Let $\Gamma \vdash M:A$ be a term of \sci, where $\Gamma$ contains
  only $\Var$-typed variables. We say that $M$ is \emph{\scim-expressive} iff:
  \begin{enumerate}[$\bullet$]
  \item $A$ is a ground type and there exists a term $\Gamma \vdash
    M':A$ of \scim\ such that $\sem{M} = \sem{M'}$ and for all stores
    $\sigma$ and values $\Gamma \vdash V:A$
\[
  \sigma, M \eval \sigma', V \iff  \sigma, M' \eval \sigma', V
\]
  or
\item $A = A_1 \llto A_2$ is a function type and for all
  \scim-expressive terms $\Delta \vdash N: A_1$, $\Gamma, \Delta
  \vdash MN: A_2$ is \scim-expressive. 
  \end{enumerate}
\end{definition}
Note that the first case above implies that all ground-type terms of
\scim\ with only $\Var$-typed free variables are automatically
\scim-expressive.

\begin{lem}
\label{lem:scim-expressive}
  Let $x_1:A_1, \ldots, x_n:A_n \vdash M:A$ be any term of \sci, and
  let $\Gamma_i \vdash N_i:A_i$ be \scim-expressive terms. Then 
  $M[\vec{N_i}/\vec{x_i}]$ is \scim-expressive. 
\end{lem}
\proof
  By induction on the structure of $M$. 
\begin{enumerate}[\hbox to8 pt{\hfill}]
\item\noindent{\hskip-12 pt\bf Variables:}\ this case is trivial.
  \item\noindent{\hskip-12 pt\bf Constants:}\ trivial since constant terms are themselves
    \scim-terms. 
  \item\noindent{\hskip-12 pt\bf Term formers of \scim:}\ for terms such as $\while{M_1}{M_2}$, we
    must prove that
    $\while{M_1[\vec{N}/\vec{x}]}{M_2[\vec{N}/\vec{x}]}$ is
    \scim-expressive.

   The subterms $M_i[\vec{N}/\vec{x}]$ are \scim-expressive by
   inductive hypothesis, and hence there are terms $M'_1$ and $M'_2$
   of \scim\ such that
\[
  \sem{M'_i} = \sem{M_i[\vec{N}/\vec{x}]}
\]
   for $i = 1, 2$, and for all stores $\sigma$ and values $V$, 
\[
  \sigma, M'_i \eval \sigma', V \iff
\sigma, M_i[\vec{N}/\vec{x}] \eval \sigma', V. 
\]
   By the definition of the operational semantics it follows that
\[
 \sigma, \while{M'_1}{M'_2} \eval \sigma', V
\]
  if and only if
\[
 \sigma, \while{M_1[\vec{N}/\vec{x}]}{M_2[\vec{N}/\vec{x}]} \eval
 \sigma', V. 
\]
  By the compositionality of the denotational semantics,
\[
 \sem{\while{M'_1}{M'_2}} = 
\sem{\while{M_1[\vec{N}/\vec{x}]}{M_2[\vec{N}/\vec{x}]}}
 \]
  and hence $\while{M_1[\vec{N}/\vec{x}]}{M_2[\vec{N}/\vec{x}]}$ is
  \scim-expressive, as required. 

  The cases of other term-formers which are included in $\scim$, such
  as $\mathtt{if}$ and  $\mathtt{new}$, are similar.

\item\noindent{\hskip-12 pt\bf Abstraction:}\ For a term $\lambda x. M$, we must prove that
 $\lambda x. M[\vec{N}/\vec{x}]$ is \scim-expressive. Let us write $M'$ for 
 $M[\vec{N}/\vec{x}]$.  By the definition of
 \scim-expressive, we must show that for all \scim-expressive terms
 $P_1$, \ldots, $P_k$ such that $(\lambda x. M') P_1 \ldots P_k$
 is of ground type, $(\lambda x. M') P_1 \ldots P_k$ is
 \scim-expressive. 

  By the inductive hypothesis, $M'[N/x]$ is \scim-expressive whenever
  $N$ is. Hence by definition of \scim-expressivity, 
  $M'[P_1/x] P_2 \ldots P_k$ is \scim-expressive whenever the $P_i$
  are. Therefore there is a term $M''$ of \scim\ such that 
  $\sem{M''} = \sem{M'[P_1/x] P_2 \ldots P_k}$ and for all stores
  $\sigma$ and values $V$, 
\[
 \sigma, M'' \eval \sigma', V \iff
\sigma, M'[P_1/x] P_2 \ldots P_k \eval \sigma', V. 
\]
But by soundness of $\beta$-reduction, 
 \[
 \sem{(\lambda x. M') P_1 \ldots P_k} = 
 \sem{M'[P_1/x] P_2 \ldots P_k} = \sem{M''}. 
 \]
This is to say that $(\lambda x. M') P_1 \ldots P_k$ is
\scim-expressive whenever the $P_i$ are, so $\lambda x.M'$ is
\scim-expressive. 
\item\noindent{\hskip-12 pt\bf Application:}\ For a term $M_1M_2$, we must show that
 $M_1[\vec{N}/\vec{x}] M_2[\vec{N}/\vec{x}]$ is \scim-expressive. But
 by inductive hypothesis, 
\[
 M_i[\vec{N}/\vec{x}]
\]
is \scim-expressive for $i=1, 2$ and the result follows by definition
of \scim-expressivity at function types. 
  \end{enumerate}
\qed

\begin{lem}
\label{lem:sound}
  For any closed term $M$ of type $\Nat$ or $\Comm$, $M \eval V$ iff $\sem{M} =
  \sem{V}$. 
\end{lem}
\proof
  By Lemma~\ref{lem:scim-expressive}, $M$ is \scim-expressive and
  hence there is a 
  term $M'$ of \scim\ such that $\sem{M} = \sem{M'}$ and $M \eval V$
  if and only if $M' \eval V$. By the soundness for \scim-terms,
  Corollary~\ref{cor:scim-sound},
  $M'\eval V$ if and only if $\sem{M'} = \sem{V}$, and the result
  follows. 
\qed

\begin{thm}[Equational Soundness]\label{thm:soundness}
  If $\Gamma \vdash M, N: A$ are terms such that $\sem{M} = \sem{N}$,
  then $M$ and $N$ are contextually equivalent. 
\end{thm}
\proof
  Since the semantics is compositional, for any context $C[-]$, we
  have $\sem{C[M]} = \sem{C[N]}$. By Lemma~\ref{lem:sound}, $C[M]\eval
  V$ iff $\sem{C[M]} = \sem{V}$ iff 
  $\sem{C[N]} = \sem{V}$ iff $C[N] \eval V$ as required. 
\qed

\section{Two extensions to the language}
\label{sec:extensions}

\noindent In the next section it will be useful to consider a version of \sci\
extended with two new constructs: erratic choice and a ``bad
variable'' constructor. It will turn out that in a certain sense these
extensions add no new expressive power---in technical parlance, they
are \emph{conservative} extensions---but they do alter the character
of the language at an intuitive level, and allow new programs to be
written. More importantly for our purposes, they give rise to the
presence of a \emph{universal type} in the language.

\subsection{Erratic choice}
\label{sec:erratic-choice}

There are several ways to add an erratic choice operation to the
language. As long as we are interested only in the ``may-converge''
version of the $\eval$ predicate, recording what values are possible
as the result of a computation without making any guarantee of
termination, the simplest form of erratic choice is a random number
generator. 

We add to the language a constant $\random$, with typing rule
\[
\begin{prooftree}
  \justifies
  \Gamma \vdash \random: \Nat
\end{prooftree}
\]
and operational semantics
\[
 \begin{prooftree}
  \justifies
  \sigma, \random \eval \sigma, n
\end{prooftree}
\]
for any $n$. 

The denotational semantics of $\random$ in our model is 
\[
 \sem{\Gamma \vdash \random: \Nat} = 
 \{ (\vec{\emptyseq}, n) \mid n \in \naturals \}. 
\]

\subsubsection{Remark}
Note that if we were to treat the must-converge predicate, this
unbounded nondeterminism would be very different from finite
nondeterminism, and would lead to some technical difficulties in the
semantics, cf.~\cite{PlotkinGD:counra}. However, for may-convergence,
adding $\random$ to the language is equivalent to adding a mere binary
nondeterministic choice operator.

\subsection{Bad variable constructor}
\label{sec:mkvar}

We alluded earlier to the ``object-oriented'' nature of our
denotational semantics of the $\Var$ type: $\Var$ is seen as the
product of countably many assignment methods of type $\Comm$ and a
dereferencing method of type $\Nat$. We can import this reading of the
$\Var$ type into the syntax of the language by means of a bad-variable
constructor $\mathtt{mkvar}$, as follows. 

The typing rule is
\[
\begin{prooftree}
  \Gamma \vdash M: \Nat \llto \Comm \quad \Gamma \vdash N : \Nat
\justifies
  \Gamma \vdash \mkvar{M}{N}: \Var
\end{prooftree}
\]
For operational semantics, there are three rules:
\[
\begin{prooftree}
  \justifies
  \sigma, \mkvar{M}{N} \eval  \sigma, \mkvar{M}{N}
\end{prooftree}
\]
\[
\begin{prooftree}
   \sigma, N \eval \sigma', n \quad
   \sigma', M \eval \sigma'', \mkvar{M_1}{M_2} \quad
   \sigma'', M_1 n \eval \sigma''', V
  \justifies
  \sigma, M \assign N \eval \sigma''', V
\end{prooftree}
\]
\[
\begin{prooftree}
  \sigma, M \eval \sigma', \mkvar{M_1}{M_2} \quad
  \sigma', M_2 \eval \sigma'', V
  \justifies
  \sigma, \deref M \eval  \sigma'', V
\end{prooftree}
\]
The idea is that $\mkvar{M}{N}$ is a variable for which the
assignment methods are given by the $Mn$ and the dereferencing method
is given by $N$. Thus any genuine variable $x$ is equivalent to 
\[
 \mkvar{(\lambda n. x \assign n)}{(\deref x)}
\]
but many other kinds of variable are available, some with very
un-variable-like behaviour, such as
\[
\mkvar{(\lambda n. \Skip)}{(3)}
\]
which always gives $3$ when dereferenced. 

%\textbf{TODO: more abstract semantics}
% DONE! 

The denotational semantics of $\mathtt{mkvar}$ is as follows.
\[
 \sem{\mkvar{M}{N}} = 
 \{ (\vec{s}, \writemove(n)) \mid (\vec{s}, *) \in \sem{Mn}
 \}
 \cup
 \{ (\vec{s}, \readmove(n)) \mid (\vec{s}, n) \in \sem{N}
 \}
\]
A somewhat more abstract presentation can be given. First note that
the denotations of terms
\[ 
 f: \Nat \llto \Comm \vdash f n : \Comm
\]
for each $n$ give us $\omega$-many maps $\sem{\Nat \llto \Comm}
\rightarrow \sem{\Comm}$ and thus a map 
\[
 \mathsf{flatten}: \sem{\Nat \llto \Comm} \rightarrow \sem{\Comm}^\omega
\]
which ``flattens'' a function into a tuple. Since $\sem{\Var} =
\sem{\Comm}^\omega \times \sem{\Nat}$ we can then define
\[
 \sem{\mkvar{M}{N}} = \pair{\sem{M}; \mathsf{flatten}}{\sem{N}}.
\]

\subsubsection{Remark}
One might argue that the $\mathtt{mkvar}$ constructor is unnatural from a
programmer's point of view. However, the ability to define one's own
assignment and dereferencing operators is a useful programming
technique which is frequently exploited in languages such as Ruby, for
example~\cite{FlanaganD:rubpl}. This constructor appears in the syntax of
most Algol-like languages which have been studied in the theoretical
literature, and is available in most models of such languages too. Our
result, to follow, which shows that $\mathtt{mkvar}$ is a conservative
extension of \sci\ is therefore somewhat comforting; moreover this
result can be extended to full Idealized Algol, arguing via a
game-based model~\cite{McCuskerGA:onsbv}. 

%\textbf{TODO: update this remark and talk about ruby}
% DONE

\subsubsection{Terminology}
We shall refer to the language $\sci$ extended with $\mathtt{mkvar}$
as $\scimk$. The relation of contextual equivalence for this language,
defined in the same way as for $\sci$, will be denoted
$\cequivmk$. Note that this relation may distinguish more terms of the
pure $\sci$ language than does $\cequiv$, because contexts may now
make use of $\mathtt{mkvar}$; in fact we shall see later that this is
not the case, so that $\mathtt{mkvar}$ is a \emph{conservative
  extension} of the language. Similarly, the language extended with both
$\mathtt{mkvar}$ and $\random$ will be called $\scicm$ and its notion
of contextual equivalence will be written $\cequivcm$.

\subsection{Soundness}
\label{sec:soundness-mkvar-choice}
We now show that the model of the extended language $\scicm$ is
sound. The proof is a straightforward extension of the arguments used
to establish 
Lemma~\ref{lem:sound}. For the sake of completeness (of the paper, not
the model!) we give the formulation here. 
\begin{definition}
  A term $x_1: \Var, \ldots, x_n: \Var \vdash M: A$ of 
  $\scicm$ is \emph{good} iff
  \begin{enumerate}[$\bullet$]
  \item $A$ is $\Comm$ and for all $\sigma$, $\sigma'$, 
\[
  \sigma, M \eval \sigma', \Skip
\]
 if and only if
 \[
 \exists (\vec{s}, *) \in \sem{M}. \strans{\sigma}{\vec{s}}{\sigma'}. 
\]
\item  $A$ is $\Nat$ and for all $\sigma$, $\sigma'$, $n$, 
\[
  \sigma, M \eval \sigma', n
\]
 if and only if
 \[
 \exists (\vec{s}, n) \in \sem{M}. \strans{\sigma}{\vec{s}}{\sigma'}. 
\]
\item $A$ is $\Var$ and for all $n$, $M \assign n$ is good and 
$\deref M$ is good. 
\item $A$ is $A_1 \llto A_2$ and for all good $N: A_1$, $MN:A_2$ is
  good. 
  \end{enumerate}
\end{definition}

\begin{lem}
  \label{lem:scicm-good}
  For any term $x_1:A_1, \ldots, x_n:A_n \vdash M:B$ of $\scicm$, if
  $\Gamma_i \vdash M_i: A_i$ are good terms for $i = 1, \ldots, n$, 
   with the $\Gamma_i$ disjoint, then
  $\Gamma_1, \ldots, \Gamma_n \vdash M[\vec{M_i}/\vec{x_i}]: B $ is
  good. 
\end{lem}
\proof
  By induction on the structure of $M$. We treat only the cases of
  $\random$ and $\mathtt{mkvar}$; the arguments for the others are as
  in the proofs of Lemmas~\ref{lem:scim-good}
  and~\ref{lem:scim-expressive}. 

  For $\random$, the operational semantics says that
\[
 \sigma, \random \eval \sigma, n
\]
 for any $\sigma$ and $n$. But
 $\strans{\sigma}{\vec{\emptyseq}}{\sigma}$ and 
\[
 (\vec{\emptyseq}, n) \in \sem{\random}
\]
 by definition. Conversely, if
 $\strans{\sigma}{\vec{\emptyseq}}{\sigma'}$ then $\sigma = \sigma'$,
 so both directions of the required implication hold. 

 For $\mathtt{mkvar}$, we shall show that if $M: \Nat \llto \Comm$ and
 $N: \Nat$ are good, then so is $\mkvar{M}{N}$.

 We must show that $(\mkvar{M}{N}) \assign n$ and $\deref (\mkvar{M}{N})$
 are good. By the definition of the operational semantics, 
\[
 \sigma, (\mkvar{M}{N}) \assign n \eval \sigma', \Skip 
\]
 if and only if
\[
 \sigma, Mn \eval \sigma', \Skip. 
\]
 Since $M$ and $n$ are good, this happens if and only if
\[
 \exists (\vec{s}, *) \in
 \sem{Mn}. \strans{\sigma}{\vec{s}}{\sigma'}. 
\]
 By definition of the semantics of $\mathtt{mkvar}$, this holds iff
\[
 \exists (\vec{s}, \writemove(n)) \in
 \sem{\mkvar{M}{N}}.\strans{\sigma}{\vec{s}}{\sigma'}
\]
 which in turn holds iff
\[
 \exists (\vec{s}, *) \in
 \sem{(\mkvar{M}{N})\assign n}.\strans{\sigma}{\vec{s}}{\sigma'}
\]
 by definition of the semantics of assignment, which completes the
 argument. The case for dereferencing is proved similarly. 
\qed

\begin{cor}
  \label{cor:scicm-sound}
  For any closed term $M$ of $\scicm$ having type $\Comm$, 
$ M \eval \Skip \Leftrightarrow * \in \sem{M}, $
 and for any closed term $M$ of type $\Nat$, 
$ M \eval n \Leftrightarrow n \in \sem{M}. $\qed
\end{cor}
Note that the statement of this result is a little different from the
analogous result for \sci, Corollary~\ref{cor:scim-sound}, because of
the nondeterminism in the language. 

Just as before, this result is enough to allow us to establish the
soundness of our model. 

\begin{thm}
  \label{thm:scicm-soundness}
 If $M$ and $N$ are terms of $\scicm$ of the same type and $\sem{M} =
 \sem{N}$, then $M \cequivcm N$. 
\end{thm}
Another simple corollary will prove useful for us later. 
\begin{cor}
\label{cor:base-type-fa}
  If $M$ and $N$ are closed terms of $\scicm$ of type $\Nat$, then $M
  \cequivcm N \iff \sem{M} = \sem{N}$. 
\end{cor}
\proof
  The right-to-left implication is
  Theorem~\ref{thm:scicm-soundness}. Left-to-right holds because if
  $M$ and $N$ are equivalent, then $M\eval n$ if and only if $N \eval
  n$ for any $n$, so by Corollary~\ref{cor:scicm-sound}, $n \in
  \sem{M}$ if and only if $n \in \sem{N}$, that is, $\sem{M} =
  \sem{N}$. 
\qed

\section{A universal type and full abstraction}
\label{sec:universal}

\noindent We begin this section with the observation that every type-object
$\sem{A}$ in $\monrel$ is a retract of $\sem{\Nat}$, confirming our
claim that the Karoubi envelope of the monoid $\mathcal{M}$ is an
appropriate setting for modelling imperative computation. 

This would be little more than an intriguing observation but for the
fact that the maps involved in the retractions are \emph{definable} by
terms of $\scicm$. Thus, not only is $\sem{\Nat}$ a universal object for
the category of type-objects in $\monrel$, but also $\Nat$ is  a universal
\emph{type} in the language. This gives rise to a very simple proof of
the full abstraction of the model of $\scicm$. We then show that this
result restricts to the smaller language \sci\ by demonstrating that
$\scicm$ extends \sci\ conservatively. 

\begin{lem}
  Let $A$ be any countable set. The monoid $A^*$ is a retract of
  $\sem{\Nat} = \omega^*$ in $\monrel$. 
\end{lem}
\proof
  Let $f: A \rightarrow \omega$ be any injective function. We
  define maps 
\[
 \mathsf{in}: A^* \rightarrow \omega^* \quad \quad
 \mathsf{out}: \omega^* \rightarrow A^*
\]
  in $\monrel$ by the relations
  \begin{eqnarray*}
\mathsf{in} & = & \{ (a_1\cdots a_k, f(a_1)\cdots f(a_k)) \mid a_1,
\ldots, a_k \in A \}\\
\mathsf{out} & = & \{ (f(a_1)\cdots f(a_k), a_1\cdots a_k) \mid a_1,
\ldots, a_k \in A \}
  \end{eqnarray*}
It is immediately clear that these are well-defined maps in $\monrel$
and that $\mathsf{in} ; \mathsf{out} = \mathsf{id}$. 
\qed

Since every type object $\sem{A}$ is a list-monoid over a countable
set, every type-object is a retract of $\sem{\Nat}$. 

We should remark, however, that not every object used to define the semantics
of \sci\ is a retract of $\sem{\Nat}$. For example one can show that
the object $\sem{\Nat} \lltensor \sem{\Nat}$ does not have this
property. The category $\monrel$ therefore possesses some advantages
over the category $\mathcal{K}(\mathcal{M})$.

We can go further in our description of type-objects as retracts of
$\sem{\Nat}$: the retractions at hand are denotations of terms of
$\scicm$. 

\begin{definition}
  A type $A$ of SCI is a \emph{definable retract} of $\Nat$ iff there
are maps $\mathsf{in}: \sem{A} \rightarrow \omega^*$ and
$\mathsf{out}:\omega^* \rightarrow \sem{A}$ in $\monrel$ such that
$\mathsf{in} ; \mathsf{out} = \mathsf{id}_{\sem{A}}$ and furthermore
there are terms $x: A \vdash \mathtt{in}: \Nat$ and $y:\Nat \vdash
\mathtt{out}: A$ of $\scicm$ such that $\sem{\mathtt{in}} = \mathsf{in}$ and
$\sem{\mathtt{out}} = \mathsf{out}$. 
\end{definition}

\begin{thm}
  Every type of $\sci$ is a definable retract of $\Nat$. 
\end{thm}
\proof
  By induction on the structure of types. We shall give particular
  definable retractions for the types $\Nat$, $\Comm$, $\Var$ and
  $\Nat \llto \Nat$. The case of a more general function type $A \llto
  B$ is then handled inductively, by defining
\begin{eqnarray*}
  x:A \llto B \vdash \mathtt{in}_{A\llto B}: \Nat & \triangleq &
  \mathtt{in_{\Nat \llto \Nat}}(\lambda
  n:\Nat. \mathtt{in_{B}}(x(\mathtt{out}_A(n)))): \Nat \\
  y: \Nat \vdash \mathtt{out_{A \llto B}} & \triangleq & 
  \lambda a:A. \mathtt{out_B}(\mathtt{out}_{\Nat \llto
    \Nat}(y)(\mathtt{in}_A(a))): A \llto B. 
\end{eqnarray*}
  The identity maps clearly make $\Nat$ a definable retract of
  itself. For the type $\Comm$, we define
  \begin{eqnarray*}
    x: \Comm \vdash \mathtt{in}_\Comm: \Nat  & \triangleq & x; 0 \\
    y: \Nat \vdash \mathtt{out}_\Comm: \Comm & \triangleq &
    \ifzero{y}{\Skip}{\Omega}
  \end{eqnarray*}
  where $\Omega$ is any nonterminating program. It is trivial to
  verify that these terms have the required property. 

  For the type $\Var$, we make use of nondeterminism. We are going to
  encode the action of reading a value $n$ from a variable as the
  number $2n$, and writing $n$ to a variable as $2n+1$ (any effective
  encoding of a disjoint sum of naturals would do, of course). The
  $\mathtt{in}$ term randomly assigns to or dereferences from the
  variable $x$, and then returns the encoding of what it has done:
\[
  x: \Var \vdash \mathtt{in}_\Var:\Nat \triangleq
  \new{r := \random}{\mathtt{ifzero}~r\begin{array}[t]{l}
  \mathtt{then}~{2(!x)}\\
  \mathtt{else}~{(x := r-1); 2r-1}.}
  \end{array}
\]
  The semantics of $\mathtt{in}_\Var$ therefore consists of all pairs
  of the forms
\[
 ([\readmove(n)],2n) \quad\quad\mbox{and}\quad\quad
 ([\writemove(n)],2n+1).
\]
    
  The $\mathtt{out}$ term makes use  of $\mathtt{mkvar}$ to create a
  variable. Both the reading and writing parts of this variable
  evaluate the natural number $y$ once. If $y$ is of the form $2n$,
  then the variable allows $n$ to be read from it; if on the other
  hand $y$ is $2n+1$, then the variable allows $n$ to be written to
  it. No other actions are possible. 
\[
  y: \Nat \vdash \mathtt{out}_\Var:\Var \triangleq
  \mathtt{mkvar}~\begin{array}[t]{l}
  (\lambda
    n:\Nat.\mathtt{if}~y=2n+1~\mathtt{then}~\Skip~\mathtt{else}~\Omega)\\
(\new{
  z:=y}{\mathtt{if}~\mathsf{even}(!z)~\mathtt{then}~!z/2~\mathtt{else}~\Omega}). 
\end{array}
%   \mkvar{(\lambda
%     n:\Nat.\mathtt{if}~y=2n+1~\mathtt{then}~\Skip~\mathtt{else}~\Omega)}
%    {(\mathtt{if}~\mathsf{even}(y)~\mathtt{then}~y/2~\mathtt{else}~\Omega)}
\]
  The semantics of this term therefore consists of all pairs of the
  forms
\[
 ([2n],\readmove(n)) \quad\quad\mbox{and}\quad\quad
 ([2n+1],\writemove(n))
\]
  thus giving the required retraction. 

  Finally for $\Nat \llto \Nat$, the term $\mathtt{in}$ supplies the
  function with a randomly generated sequence of inputs, $s$, observes
  the output, $n$, and returns an encoding of the pair $(s,n)$ as a
  natural number. Compare this with the $\code(-)$ function used to
  embed  $[\pw \rightarrow \pw]$ in $\pw$ in Scott's model. To ease
  the notation we use a liberal dose of syntactic sugar. We assume
  that an encoding of sequences of natural numbers as naturals exists,
  and suppress mention of it, so it appears that the
  variable $s$ in the term below is used to store finite sequences
  directly. We write $\emptyseq$ for the encoding of
  the empty sequence, $[n]$ for the encoding of the singleton sequence
  containing the element $n$, and $\cdot$ for the encoding of
  concatenation. If $n$ is a number encoding a sequence $s$,
  $\length{n}$ denotes the length of sequence $s$ and $n_i$ denoting
  the $i$th element of $s$. We also use pair notation $\langle
  s,n\rangle$ for the encoding of this pair as a natural
  number, and $\mathtt{fst}$ and $\mathtt{snd}$ to compute the
  projections from such encoded pairs. Finally we allow multiple
  variables to be allocated and 
  initialized at once, so that $\mathtt{new}~{s:=\emptyseq; x
    :=0}~\mathtt{in}~M$ means $\new{s}{\new{x}{s:=\emptyseq; x:=0;
      M}}$. With these abbreviations at our disposal,
  $\mathtt{in}_{\Nat\llto\Nat}$ is defined as follows. 

\[
 f: \Nat \llto \Nat \vdash \mathtt{in}_{\Nat \llto \Nat} \triangleq
   \mathtt{new}~\begin{array}[t]{l}
      {s:=\emptyseq; x :=0}~\mathtt{in}\\
     {x :=  f(\new{r:=\random}{(s:= !s\cdot [!r]);  !r});}\\ 
     \langle !s, !x\rangle. 
   \end{array}
\]
  Finally for $\mathtt{out}_{\Nat \llto \Nat}$, we take the value $y:\Nat$, 
  decode it as a pair $(s,n)$, and return a function which can return
  $n$ on observation of the input sequence $s$, but can do nothing
  else.
\[
  y:\Nat \vdash \mathtt{out}_{\Nat \llto \Nat} \triangleq
    \lambda z^\Nat. \mathtt{new}~
\begin{array}[t]{l}
{y':= y; z':= z; s:=\mathtt{fst}(!y'); n:=
      \mathtt{snd}(!y'); x:=0}~\mathtt{in}\\
\mathtt{while}~\begin{array}[t]{l}
        {!x < |!s|}~\mathtt{do}\\
        {\mathtt{if}~!z'_{!x}={!s}_{!x}~\mathtt{then}~x:=!x+1~\mathtt{else}~\Omega;}
              \end{array}\\
  !n 
            \end{array}
\]
\qed

These definable retractions allow us to prove full abstraction for
$\scicm$ in a very straightforward fashion. 
\begin{thm}
  \label{thm:scicm-fa}
  The model of $\scicm$ in $\monrel$ is fully abstract. That is, for
  any closed terms $M$ and $N$ of the same type, $\sem{M} = \sem{N}$
  if and only if $M \cequivcm N$. 
\end{thm}
\proof
  The left-to-right implication is~Theorem~\ref{thm:scicm-soundness}. For the
  right-to-left, suppose $M$ and $N$ are equivalent terms. Then by
  definition of equivalence, we also have
\[
 \mathtt{in}[M/x] \cequivcm \mathtt{in}[N/x]. 
\]
  These are closed terms of type $\Nat$, so by
  Corollary~\ref{cor:base-type-fa}, 
  $\sem{\mathtt{in}[M/x]}= \sem{\mathtt{in}[N/x]}$. By
  compositionality of the semantics it follows that 
  $\sem{\mathtt{out}[\mathtt{in}[M/x]/y]} 
 = \sem{\mathtt{out}[\mathtt{in}[N/x]/y]}$. But 
  $\sem{\mathtt{out}[\mathtt{in}[M/x]/y]} = \sem{M};
  \sem{\mathtt{in}}; \sem{\mathtt{out}}$ and similarly for $N$, so we
  conclude that $\sem{M} = \sem{N}$ as required. 
\qed
% However, we also know that the constructs $\mkvar$ and $\random$ are
% conservative extensions, so we have full abstraction for $\scim$ and
% $\sci$ as corollaries.
% \begin{corollary}
%   The models of $\scicm$ and $\sci$ in $\monrel$ are fully abstract.
% \end{corollary}

\section{A model without nondeterminism}
\label{sec:coherence-rel}

\noindent We have established full abstraction of our model of $\scicm$, which
admits both the $\texttt{mkvar}$ construct and nondeterminism. Before
embarking on our proof that these additional constructs do not change
the notion of equivalence in \sci, we first develop a more
constrained model in which $\random$ cannot be interpreted. 

Reddy's original object-spaces model did not admit the
nondeterministic construct $\random$. We use some of
Reddy's ideas to construct a variant of the category $\monrel$ which
contains the same model of $\scimk$ but, like Reddy's category,
contains no nondeterministic elements. The idea is to introduce a
relation of coherence, in the style of Girard's coherence
spaces~\cite{GirardJY:prot}. 

\begin{definition}
  Given a monoid $A$, a \emph{coherence relation} $\coh$ on $A$ is a
  symmetric reflexive binary relation on the underlying set of $A$ such that
\begin{enumerate}[\hbox to8 pt{\hfill}]
\item\noindent{\hskip-12 pt\bf prefix closure:}\ if $a_1a_2 \coh a'_1a'_2$ then $a_1 \coh a'_1$
  \item\noindent{\hskip-12 pt\bf extension:}\ if $aa_1 \coh aa_2$ then $a_1 \coh a_2$. 
\end{enumerate}
\end{definition}

A useful intution is that elements $a$ and $a'$ are coherent, $a\coh
a'$, if they can coexist as possible observations to be made of a
single deterministic computation at the same state. 
So, for instance, distinct natural numbers $n$ and $n'$
will not be coherent in the denotation of $\Nat$, but $\writemove(n)$ and
$\writemove(n')$ will be coherent in $\Var$ because a variable may
allow any value to be written to it. 

\begin{definition}
  The category $\monrelcoh$ is defined as follows. Objects are pairs
  $(A, \coh)$ consisting of a monoid $A$ together with a coherence
  relation on $A$, and maps from $(A, \coh_A)$ to $(B, \coh_B)$ are
  relations $R$ such that $R$ is a map from $A$ to $B$ in $\monrel$
  and furthermore
  \begin{enumerate}[$\bullet$]
  \item if $a \coh_A a'$, $aRb$ and $a'Rb'$ then $b\coh_B b'$
  \item if $a \coh_A a'$, $a R b$ and $a'Rb$ then $a = a'$. 
  \end{enumerate}
  Composition is the usual composition of relations. 
\end{definition}

\begin{lem}
  $\monrelcoh$ is a category. 
\end{lem}
\proof
  It is clear that the identity relations are valid maps in
  $\monrelcoh$ so we just need to show that composition preserves the
  two new constraints on maps. Let $R: A\rightarrow B$ and
  $S:B\rightarrow C$ be maps in $\monrelcoh$. Suppose $a \coh_A a'$
  and that $a R ; S c$ and $a' R; S c'$. Then there exist $b, b' \in
  B$ such that $aRb$, $bS c$, $a'R b'$ and $b'S c'$. Since $a \coh_A
  a'$ we have $b \coh_B b'$ and hence $c \coh_C c'$ as required. Now
  suppose $c=c'$; we shall show that $a = a'$. Since $S$ is a valid
  map, we have $b = b'$ and then since $R$ is valid, $a = a'$. Hence
  $R;S$ is a valid map in $\monrelcoh$. 
\qed

The following definition is due to Reddy~\cite{ReddyUS:gloscu}. 
\begin{definition}
  Given a set $A$ and a symmetric reflexive binary relation $\coh_A$ on $A$, we
  define an object of $\monrelcoh$ called the 
  \emph{object-space over $A$} consisting of the 
  free monoid over $A$ with coherence relation defined by:
\[
 a_1\ldots a_m \coh a'_1\ldots a'_n 
\]
if and only if 
\[
 \forall i\in\{1, \ldots \mathsf{min}(m,n) -1\}. a_1\ldots a_i =
 a'_1\ldots a'_i \Rightarrow a_{i+1}\coh_A a'_{i+1}. 
\]  
 That is to say, two sequences are coherent if either one is a prefix
 of the other, or at the first place they differ, the two differing
 elements are coherent. 
\end{definition}

\begin{lem}\label{lem:monrelcoh-free}
  Let $(A, \coh)$ be a set with a coherence relation, and let $A^*$ be
  the object-space over this structure. Let $B$ be any object of
  $\monrelcoh$. Let $R$ be a relation from $UB$ to $A$ such that if $b
  R a$ and $b' R a'$ with $b \coh b'$ then $a \coh a'$ and if $a =
  a'$ then $b = b'$. Then there is a unique map in $\monrelcoh$ from
  $B$ to $A^*$ which extends $R$; by abuse of notation we also write
  $R$ for this relation.
\end{lem}
\proof
  The unique candidate for this map is the extension of $R$ to a map
  $B$ to $A^*$ in $\monrel$, exploiting the fact that $A^*$ is the
  free monoid over $A$. We just need to show that it is a valid map in
  $\monrelcoh$. 

  We first show that if $b \coh b'$ with $b R a_1\cdots a_n$ and $b' R
  a'_1 \cdots a'_{n'}$ then $a_1\cdots a_n \coh a'_1\cdots
  a'_{n'}$. This requires demonstrating that at the first $i$ such
  that $a_i \not =  a'_i$, we have $a_i \coh a'_i$, if such an $i$
  exists. We proceed by induction on the minimum of $n, n'$. In the
  base case there is nothing to prove, so suppose both $n$ and $n'$
  are non-zero. 

  By the decomposition property, we can find $b_1, \ldots, b_n$ such
  that $b = b_1\cdots b_n$ and each $b_i R a_i$, and similarly for
  $b'$ and the $a'_i$. By the prefix-closure property in $B$, $b_1
  \coh b'_1$ and hence $a_1 \coh a'_1$. Thus if $a_1 \not = a'_1$, we
  are done. Otherwise, $a_1 = a'_1$ implies that $b_1 = b'_1$ and then
  by the extension property of coherence in $B$, we have $b_2\cdots
  b_n \coh b'_2 \cdots b'_{n'}$ and of course $b_2\cdots b_n R
  a_2\cdots a_n$ and similarly for the $b'_i$ and $a'_i$. Then the
  inductive hypothesis gives us the result we require. 

  We now show that if additionally $a_1\cdots a_n = a'_1 \cdots a'_{n'}$
  then $b = b'$, again by induction on $n$ (which is equal to
  $n'$). The base case is guaranteed by the identity reflection
  property of maps in $\monrel$. In the inductive step, we again
  decompose $b$ and $b'$ as above, and note that since $a_1 = a'_1$ we
  have $b_1 = b'_1$. Then we also have $b_2\cdots b_n R a_2\cdots
  a_n$ and similarly for the $b'_i$, and conclude by the inductive
  hypothesis. 
\qed

The product, tensor and exponential constructions in $\monrel$ all
lift to $\monrelcoh$. This can be expressed as follows. 

\begin{lem}
  $\monrelcoh$ is a symmetric monoidal category with products, and the
  object-spaces form an exponential ideal in $\monrelcoh$. Moreover
  the forgetful functor to $\monrel$ preserves all this structure on
  the nose. 
\end{lem}
\proof
  We just need to define the coherence-relation parts of the various
  constructions and show that they are well-defined and have the
  appropriate properties. 

  For the monoidal structure, coherence is defined pointwise:
\[
 (a,b) \coh_{A \lltensor B} (a', b') \iff a \coh_A a, b \coh_B b. 
\]
  (To aid legibility in future we will drop the subscripts on the
  $\coh$ relations where no confusion will arise.)

  It is clear that this definition makes $\lltensor$ into a bifunctor
  on $\monrelcoh$ and that the associativity, symmetry and unit maps
  from $\monrel$ are well-defined maps in $\monrelcoh$ too. 

  We now consider the exponentials. Let $(A, \coh_A)$ be an object of
  $\monrelcoh$, and let $(B, \coh_B)$ be a set equipped with a
  symmetric reflexive binary
  relation. In $\monrel$ the exponential $A \llto B^*$ is given by
  the free monoid over $UA \times B$. We shall define a symmetric reflexive
  binary relation on this set and show that the object-space this
  defines is the required exponential in $\monrelcoh$. 

  The coherence relation on $UA \times B$ echoes the definition of map
  in $\monrelcoh$: $(a, b) \coh (a',b')$ if and only if 
  \begin{enumerate}[$\bullet$]
  \item $a\coh_A a' \Rightarrow b \coh_B b'$
  \item $a \coh_A a' \land b = b' \Rightarrow a = a'$. 
  \end{enumerate}

  By Lemma~\ref{lem:monrelcoh-free}, maps from an object $C$ into this
  object space are described by relations from $UC$ to $UA \times B$
  which satisfy the appropriate coherence constraints. That is, if 
  $c R (a,b)$ and $c' R (a',b')$ then we have 
  \begin{enumerate}[$\bullet$]
  \item $c \coh_C c' \implies (a,b) \coh (a',b')$ 
  \item $c \coh_C c' \land (a,b) = (a',b') \implies c = c'$. 
  \end{enumerate}
  
  On the other hand, maps from $C \lltensor A$ to $B^*$ are given by
  relations from $UC \times UA$ to $B$ such that
  \begin{enumerate}[$\bullet$]
  \item $c \coh_C c' \land a \coh_A a' \implies b \coh_B b'$
  \item $c \coh_C c' \land a \coh_A a' \land b = b' \implies a = a' \land
    c = c'$. 
  \end{enumerate}
  
  It is straightforward to verify that these are the same constraints,
  so that we have a natural bijection of homsets: 
\[
\monrelcoh(C \lltensor A, B^*) \cong \monrelcoh(C, A \llto B),\]
 as required. 

  A similar argument shows that products in $\monrel$ lift to
  $\monrelcoh$. For object-spaces, the construction is very
  straightforward: the product of object-spaces $A^*$ and $B^*$ is the
  object space over the disjoint union $A + B$, equipped with the
  coherence relation which relates elements of $A$ if and only if they
  are related in the object space $A^*$, and similarly for $B$, but
  also relates all elements of $A$ to all elements of $B$. 
\qed

$\monrelcoh$ therefore possesses all the structure we require to model
$\sci$. To lift our model to $\monrelcoh$ we just need to give
interpretations of the base types and constants. The base types are
all interpreted using object spaces, with underlying coherence
relations as follows:
\begin{enumerate}[$\bullet$]
\item for $\Nat$, $n \coh n' \iff n = n'$.
\item for $\Comm$, $* \coh *$.
\item for $\Var$, $\writemove(n) \coh \writemove(n')$ for all $n, n'$;
   $\readmove(n) \coh \readmove(n') \iff n = n'$; and $\writemove(n)
   \coh \readmove(n')$ for all $n, n'$. Note that this makes $\Var $
   the product object-space of $\Nat$ with $\omega$-many copies of
   $\Comm$. 
\end{enumerate}
It is easy to check that the constant maps used in the
denotations of $\sci$ terms are maps of $\monrelcoh$ over the
appropriate types. The same applies
to $\mathtt{mkvar}$, but not to $\random$: the map $\sem{\random}$
clearly violates the coherence constraints since it returns incoherent
outputs from coherent (empty) inputs. 

\begin{thm}
The model of $\scimk$ in $\monrel$ lifts to $\monrelcoh$. \qed
\end{thm}

\begin{cor}
\label{cor:coherence}
  If $\vdash M: A$ is a closed term of $\scimk$ and $a, a' \in \sem{M}$ then 
  $a \coh a'$. (Here we blur the distinction between maps from the
  tensor unit into $\sem{A}$ and subsets of $\sem{A}$.)\qed
\end{cor}
Thus the model of $\scimk$ in $\monrelcoh$ captures $\scimk$'s
deterministic nature: for instance, closed terms of type $\Nat$ contain
at most one natural number in their denotation. 

\section{Conservativity results}
\label{sec:conservativity}

\noindent In this section we show that the extensions of $\sci$ with the
$\mathtt{mkvar}$ and $\random$ operators are \emph{conservative}, that
is to say, they have no effect on the relation of contextual
equivalence for terms of the original $\sci$ language. This means that
the new contexts available when the language is extended have no
additional discriminating power, and as a result, the full abstraction
theorem for $\scicm$ also applies to the smaller languages $\scimk$
and $\sci$. As explained in~\cite{McCuskerGA:fularm}, this work shows
that Reddy's object-spaces model~\cite{ReddyUS:gloscu} was the first
example of a fully abstract semantics for a higher-order imperative
language, though this was not known at the time. Its full abstraction
is remarkable since it contains a great many undefinable
elements. However, the definable elements do suffice to distinguish
any two different elements of the model, and it is this which leads to
full abstraction.

Though we present our results in the form of conservativity theorems
rather than direct full abstraction proofs, our arguments hinge on
partial definability results which would be enough to establish full
abstraction of the model for $\sci$ and $\scimk$ directly, that is,
without appealing to Theorem~\ref{thm:scicm-fa}, if desired. The proof
of conservativity of 
$\mathtt{mkvar}$  in particular makes heavy use of our definability
results, and is essentially the same as the direct proof of full
abstraction given in~\cite{McCuskerGA:fularm}. Nevertheless we believe
that presenting the results as conservativity theorems is worthwhile,
particularly in light of the relatively cheap proof of full
abstraction for $\scicm$, and the limited use of definability in the
proof of conservativity of $\random$.

\subsection{Definability}
\label{sec:definability}

As explained above, our conservativity results are established by
means of a partial definability result which demonstrates how certain
elements of our model are found as the denotations of terms from
$\sci$ and its extensions.

Let us first mention a curious fact. Let $C[-]$ be some context
of $\sci$, so that in particular $C[-]$ does not employ
$\mathtt{mkvar}$. If 
\[
 C[\mathtt{if}~!x=3~\mathtt{then~skip~else~diverge}]\cvgs, 
\]
then it is also the case that $ C[x:=3]\cvgs.$
This inability of $\mathtt{mkvar}$-free
contexts to distinguish completely between reading and writing into
variables is the main obstacle to overcome in our definability
proof. The presence of $\mathtt{mkvar}$ makes quite a difference,
since for example a context binding $\mathtt{x}$ to the term
\[
 \mkvar{(\lambda y.\mathtt{diverge})}{(3)}
\]
will make the first term above converge and the second diverge. This
immediately tells us that the addition of $\mathtt{mkvar}$ is not
conservative with respect to the contextual \emph{preorder}. Our work
in this section will show that it is conservative with respect to
contextual \emph{equivalence}; this came as a surprise. 

The following definition captures the relationship between sequences
of observations which is at work in the above example. 
\begin{definition}
  For any SCI type $A$, we define the \emph{positive and negative
    read-write orders} 
  $\rworder^+$ and $\rworder^-$ between elements of $\sem{A}$ as
  follows. We give only the definitions for singleton elements; the
  definitions are extended to sequences by requiring that the elements
  of the sequences are related pointwise. 
  \begin{enumerate}[$\bullet$]
  \item At type $\Comm$: 
\[
  * \rworder^+ * \land * \rworder^- *
\]
\item At type $\Nat$: 
\[
 n \rworder^+ m \iff n = m \iff n \rworder^- m
\]
  \item At type $\Var$: 
\[
\begin{array}{rclcl}
 a & \rworder^+ & a' & \iff & (a = a') \lor \exists n. a = \readmove(n)
              \land a' = \writemove(n) \\
 a & \rworder^- & a' & \iff & a = a'
\end{array}
\]
\item At type $A \llto B$: 
\[
\begin{array}{rclcl}
 (s,b) & \rworder^+ & (s',b') & \iff & s \rworder^- s' \land b
 \rworder^+ b' \\
 (s,b) & \rworder^- & (s',b') & \iff & s \rworder^+ s' \land b
 \rworder^- b'
\end{array}
\]
  \end{enumerate}
In general, $s \rworder^+ t$ iff $t$ can be obtained from $s$ by
replacing some occurrences of $\readmove(n)$ actions in positive
occurrences of the type $\Var$ by the corresponding $\writemove(n)$
actions. The order $\rworder^-$ is the same but operates on actions in
negative occurrences of $\Var$. 
\end{definition}

We are now in a position to state our definability result. 
\begin{lem}\label{lem:definability}
  Let $A$ be any type of $\sci$\ and let $a \in \sem{A}$ be any
  element of the monoid interpreting $A$. There exists a term 
\[
x:A \vdash \test(a):\Comm
\]
  of $\sci$ (not including $\mathtt{mkvar}$ or $\random$) such that
  $(s,*) \in \sem{\test(a)}$ iff $a \rworder^- s$.  
  There also exists a
  context $\Gamma = x_1: \Var, \ldots, x_n: \Var$, $\Gamma$-stores
  $\init{a}$ and $\final{a}$, and a term 
\[
 \Gamma \vdash \produce(a):A
\]
 such that for all $a' \in \sem{A}$, 
\[
(\exists s. (s,a') \in \sem{\produce(a)} \land
\strans{\init{a}}{s}{\final{a}})
\iff 
a \rworder^+ a'. 
\]
% there exists 
%  $(s,a') \in \sem{\produce(a)}$ with $\strans{\init{a}}{s}{\final{a}}$
%  if and only if $a \rworder^+ a'$.  
\end{lem}
\proof
  We will prove the two parts of this lemma simultaneously by
  induction on the type $A$. First note that any $a \in \sem{A}$ is a
  sequence of elements from a certain alphabet. Before beginning the
  main induction, we show that it suffices to consider the case when
  $a$ is a singleton sequence. The cases when $a$ is empty are
  trivial: $\test([]) = \Skip$ and $\produce([])$ is any divergent
  term, with $\init{[]}$ and $\final{[]}$ both being the unique store
  on no variables. 

If $a = [a_1,a_2,\ldots,a_n]$, then we can define $\test(a)$ as 
\[
 \test([a_1])\comp \test([a_2]) \comp \ldots \comp \test([a_n]). 
\]
  For the $\produce$ part, suppose that $A = A_1 \llto A_2 \llto
  \ldots \llto A_k
  \llto B$ for some base type $B$, and that the context $\Gamma$
  contains all the variables needed to define the $\produce(a_i)$. 
  For any store $\sigma$ over variables $x_1, \ldots, x_n$, 
  define $\mathsf{check}(\sigma)$ to be the term 
\[
\begin{array}{l}
 \mathtt{if}~(!x_1 \not =
 \sigma(x_1))~\mathtt{then}~\mathtt{diverge}\\
 \mathtt{else~if}~(!x_2 \not =
 \sigma(x_2))~\mathtt{then}~\mathtt{diverge}\\
 \ldots \\
 \mathtt{else~if}~(!x_n \not =
 \sigma(x_n))~\mathtt{then}~\mathtt{diverge}\\
 \mathtt{else~skip}
\end{array}
\]
 Define $\mathsf{set}(\sigma)$ to be $ x_1:=\sigma(x_1) \comp \cdots \comp x_n := \sigma(x_n).$

  An appropriate term $\produce(a)$ can then be defined as follows.
\[
 \Gamma, x:\Var \vdash \lambda \vec{y_i}^{\vec{A_i}}. 
 \begin{array}[t]{l}
   x:= !x+1 \comp \\
   \mathtt{if}~(!x = 1)~\mathtt{then}~\produce(a_1)y_1 \ldots y_k \\
   \mathtt{else~if}~(!x = 2)~\mathtt{then}~\mathsf{check}(\final{a_1})\comp\\
   \phantom{\mathtt{else~if}~(!x = 2)~\mathtt{then}~}
                                      \mathsf{set}(\init{a_2})\comp\\
   \phantom{\mathtt{else~if}~(!x = 2)~\mathtt{then}~} 
                                      \produce(a_2)y_1\ldots y_k \\
   \ldots \\
   \mathtt{else~if}~(!x = n)~\mathtt{then}~\mathsf{check}(\final{a_{n-1}})\comp\\
    \phantom{\mathtt{else~if}~(!x = n)~\mathtt{then}~}
                     \mathsf{set}(\init{a_n})\comp\\
    \phantom{\mathtt{else~if}~(!x = n)~\mathtt{then}~}
                     \produce(a_n)y_1 \ldots y_k\\
   \mathtt{else}~\mathtt{diverge}
 \end{array}
\]
The required initial state $\init{a}$ is
$\override{\init{a_1}}{x}{0}$, and the final state $\final{a}$ is
$\override{\final{a_n}}{x}{n}$. 

We now define $\test(a)$ and $\produce(a)$ for the case when $a$ is a
singleton, by induction on the structure of the type $A$. 

For the type $\Comm$, we define
\begin{eqnarray*}
  \test(*) & = & x:\Comm \vdash x:\Comm \\
  \produce(*) & = & y: \Var \vdash y := !y + 1: \Comm \\
\init{*} & = & (y \mapsto 0) \\
\final{*} & = & (y \mapsto 1)
\end{eqnarray*}
Note the way the initial and final states check that the command
$\produce(*)$ is used exactly once. 

The type $\Nat$ is handled similarly:
\begin{eqnarray*}
  \test(n) & = & x:\Nat \vdash 
 \mathtt{if}~(x = n)~\mathtt{then~skip~else~diverge}: \Comm \\
  \produce(n) & = & y: \Var \vdash y := !y + 1; n: \Nat \\
  \init{n} & = & (y \mapsto 0) \\
  \final{n} & = & (y \mapsto 1)
\end{eqnarray*}

For $\Var$, there are two kinds of action to consider: those for
reading and those for writing. For writing we define:
\begin{eqnarray*}
  \test(\writemove(n)) & = & x:\Var \vdash x:= n: \Comm \\
  \produce(\writemove(n)) & = & x: \Var, y: \Var \vdash y := !y + 1; x
  : \Var \\
  \init{\writemove(n)} & = & (x \mapsto n+1, y \mapsto 0) \\
  \final{\writemove(n)} & = & (x \mapsto n, y \mapsto 1)
\end{eqnarray*}
For $\produce(\writemove(n))$, the variable $y$ checks that exactly
one use is made, and the variable $x$ checks that the one use is a
write-action assigning $n$ to the variable. 

Reading is handled similarly:
\begin{eqnarray*}
  \test(\readmove(n)) & = & x:\Var \vdash  \mathtt{if}~(!x = n)~\mathtt{then~skip~else~diverge}: \Comm \\
  \produce(\readmove(n)) & = & x: \Var, y: \Var \vdash y := !y + 1; x: \Var \\
  \init{\readmove(n)} & = & (x \mapsto n, y \mapsto 0) \\
  \final{\readmove(n)} & = & (x \mapsto n, y \mapsto 1)
\end{eqnarray*}
In $\init{\readmove(n)}$, the variable $x$ holds $n$ so that if the
expression $\produce(\readmove(n))$ is used for a read, the value $n$
is returned. The variable $x$ must also hold $n$ finally, so
$\produce(\readmove(n))$ cannot reach the state
$\final{\readmove(n)}$ if it is  used to write a value other than
$n$. However, it would admit a single $\writemove(n)$ action. This is
the reason for introducing the $\rworder$ relation: if a term of our
language can engage in a $\readmove(n)$ action, then it can also
engage in $\writemove(n)$. 

For a function type $A \llto B$, the action we are dealing with has
the form $(s,b)$ where $s$ is a sequence of actions from $A$ and $b$
is an action from $B$. We can now define
\begin{eqnarray*}
  \test(s,b) & = & x:A \llto B \vdash 
  \begin{array}[t]{l}
    \new{x_1,\ldots,x_n}{} \\
    ~~~\mathsf{set}(\init{s}); \\
    ~~~(\lambda x^B. \test(b))(x \produce(s)); \\
    ~~~\mathsf{check}(\final{s}); 
  \end{array} \\
  \produce(s,b) & = & \lambda x^A. \test(s); \produce(b)\\
  \init{s,b} & = & \init{b} \\
\final{s,b} & = & \final{b}
\end{eqnarray*}
where $x_1, \ldots, x_n$ are the variables used in $\produce(s)$. 

The non-interference between function and argument allows us to define
these terms very simply: for $\test(s,b)$ we supply the function $x$
with an argument which will produce the sequence $s$, and check that
the output from $x$ is $b$. We must also check that the function $x$
uses its argument in the appropriate, $s$-producing way, which is done
by means of the $\init{s}$ and $\final{s}$ states. For $\produce(s,b)$
we simply test that the argument $x$ is capable of producing $s$, and
then produce~$b$. 

It is straightforward to check that these terms have the required
properties. 
\qed

\subsection{Conservativity of \texorpdfstring{$\random$}{random}}
\label{sec:conservativity-random}

\begin{lem}[$\random$ is conservative]
\label{lem:conservativity-random}
Let $\Gamma \vdash M, N: A$ be terms of $\scimk$ such that $M\cequivmk
N$. Then $M \cequivcm N$. 
\end{lem}
\proof
 It suffices to consider \emph{closed} terms, because in all
 the language fragments we consider, open terms $M$ and $N$ are equivalent
 if and only if their closures $\lambda \vec{x}.M$ and 
 $\lambda \vec{x}.N$ are equivalent. 

 So, let $\vdash M, N:A$, suppose $M \cequivmk N$ and let $C[-]$ be a
 context, possibly employing $\random$, such that $C[M]\eval
 \Skip$. We shall show that $C[N]\eval \Skip$ by induction on the
 number of occurrences of $\random$ in $C[-]$. 

 The base case, where $C[-]$ does not employ $\random$ at all, is
 trivial: $C[-]$ is a $\scimk$ context, so since $M\cequivmk N$, we
 have $C[N]\eval \Skip$. 

 For the inductive step, let $C'[-]$ be the context obtained from
 $C[-]$ by replacing one  occurrence of $\random$ with a fresh
 variable $r$ of type $\Nat$. Then for any term $P$, $C[P] \eval
 \Skip$ if and only if  $(\lambda r.C'[P])(\random) \eval \Skip$. 

 Since $(\lambda r.C'[M])(\random) \eval \Skip$,
 Corollary~\ref{cor:scicm-sound} implies that 
\[
 (\emptyseq, *) \in \sem{(\lambda r.C'[M])(\random)}.
\]
 By definition
 of $\sem{\random}$ and the semantics of application, there must exist
 a sequence $s$ of natural numbers such that 
 $(s, *) \in \sem{\lambda r.C'[M]}$. 

 By Lemma~\ref{lem:definability}, there is a term  
\[
 x: \Nat \rightarrow \Comm \vdash \mathsf{test}: \Comm
\]
not involving $\random$, such that $(t, *) \in \sem{\mathsf{test}}$
iff $ t = (s,*)$.  
 
 We therefore have $(\emptyseq, *) \in \sem{(\lambda
 x. \mathsf{test})(\lambda r.C'[M])}$ and hence by
 Corollary~\ref{cor:scicm-sound}, $(\lambda x. \mathsf{test})(\lambda
 r.C'[M]) \eval \Skip$. But $(\lambda x.\mathsf{test})(\lambda
 r.C'[-])$ is a context involving the same number of occurrences of
 $\random$ as does $C'[-]$, so by inductive hypothesis we 
 also have $(\lambda x. \mathsf{test})(\lambda r.C'[N]) \eval
 \Skip$. Therefore $(\emptyseq, *) \in \sem{(\lambda
 x. \mathsf{test})(\lambda r.C'[N])}$, which is only possible if $(s,*)
 \in \sem{\lambda r.C'[N]}$. But then 
\[
 (\emptyseq, *) \in \sem{(\lambda r.C'[N])(\random)}
\]
  and hence by
 Corollary~\ref{cor:scicm-sound} again, $(\lambda r.C'[N])(\random)
 \eval \Skip$. Finally we can conclude that $C[N]\eval \Skip$ as
 required. 
\qed

\begin{cor}
  The model of $\scimk$ in $\monrel$ is fully abstract. \qed
\end{cor}

\subsection{Conservativity of \texorpdfstring{ $\mathtt{mkvar}$}{mkvar}}
\label{sec:conservativity-mkvar}

\begin{lem}
\label{lem:coherence-and-order}
  Let $A^*$ be an object-space interpreting a type of $\sci$ in
  $\monrelcoh$ and let $a, a' \in A^*$.
  \begin{enumerate}[$\bullet$]
  \item If $a \rworder^- a'$ and $a \coh
  a'$ then $a = a'$.
\item If $a \rworder^+ a'$ then $a \coh a'$. 
  \end{enumerate}
\end{lem}
\proof
  By induction on type. We consider only the cases of singleton
  sequences; the general cases follow easily. 

  For $\Comm$ and $\Nat$, both $\rworder^-$ and
  $\rworder^+$ are the identity relations, so the results hold
  trivially. For $\Var$, $\rworder^-$ is again the identity relation
  completing this case. For $\rworder^+$, the result follows from the
  fact that $\readmove(n) \coh \writemove(n)$. 

  For the inductive step, consider elements $(s,b)$ and $(s',b')$ of
  $A \llto B$. If $(s,b) \rworder^- (s',b')$ then $s \rworder^+ s'$
  and $b \rworder^- b'$. By the inductive hypothesis on type $A$, $s
  \coh s'$ so if $(s,b) \coh (s',b')$ then we also have $b \coh
  b'$. The inductive hypothesis on $B$ then gives us $b = b'$ and
  hence $s = s'$ as required. If $(s,b) \rworder^+ (s',b')$ then $s
  \rworder^- s'$ and $b \rworder^+ b'$. Then if $s \coh s'$, the
  inductive hypothesis gives us $s = s'$. Induction also tells us that
  $b \coh b'$, and hence $(s,b) \coh (s',b')$ as required. 
\qed

\begin{lem}[$\mathtt{mkvar}$ is conservative]
\label{lem:conservativity-mkvar}
Let $\Gamma \vdash M, N: A$ be terms of $\sci$ such that $M\cequiv
N$. Then $M \cequivmk N$. 
\end{lem}
\proof
As in Lemma~\ref{lem:conservativity-random} we consider only closed
terms.
Suppose $\vdash M, N:A$ with $M
\cequiv N$ and let $(\emptyseq,a) \in \sem{M}$ be any element of the 
denotation of $M$. By~Lemma~\ref{lem:definability} there is a term
$x:A \vdash \test(a): \Comm$ such that $(a',*) \in \sem{\test(a)}$ if
and only if $a \rworder^- a'$. We therefore have $(\emptyseq, *) \in
\sem{(\lambda x. \test(a))M}$, and hence $(\lambda x.\test(a))M \eval
  \Skip$ by Corollary~\ref{cor:scim-sound}. By hypothesis we have
  $(\lambda x.\test(a))N \eval \Skip$, so that $(\emptyseq, *) \in
  \sem{(\lambda x.\test(a))N}$. Therefore there is some $a'$ such that
    $a \rworder^- a'$ and $(\emptyseq, a') \in \sem{N}$. 
  Symmetrically we can find $a''$ such that $a' \rworder^- a''$ and
  $(\emptyseq, a'') \in \sem{M}$. 

 By Corollary~\ref{cor:coherence}, $a \coh a''$ and then by
 Lemma~\ref{lem:coherence-and-order}, $a = a''$ and hence $a = a'$. It
 follows that $\sem{M} = \sem{N}$ and hence $M \cequivmk N$ by
 Theorem~\ref{thm:scicm-soundness}. 
\qed

\begin{cor}
  The model of $\sci$ in $\monrel$ is fully abstract. \qed
\end{cor}

We remark that Reddy was not aware that his model was fully abstract;
indeed it was believed not to be. 

\section{Conclusions}

\noindent We have shown that a simple amendment of Scott's $\pw$ graph-model
gives rise to a model of imperative computation, in the event-based
style of Reddy's object-spaces model and later models based on game
semantics. Moreover we have shown that this model contains a universal
type, thus yielding a very cheap proof of full abstraction for the
language $\scicm$. With some additional work we have established full
abstraction for the original \sci\ language via conservativity
results; this was not known prior to our work. 

We believe that the general approach of constructing models in this
way is of interest and has the potential to give rise to a range of
interesting concrete models and some useful insights at a more
abstract level. We intend to develop an axiomatic presentation of our
constructions, expanding on the work of Hyland et
al.~\cite{PowerJ:cattfe}. At present it is not clear whether the more
refined game-based models can be presented in this style; this remains
a topic for further investigation.

\end{document}
% Local Variables: 
% mode: latex
% TeX-master: t
% End: 